\documentclass[reprint,superscriptaddress,
 amsmath,amssymb]{revtex4-1}
\UseRawInputEncoding
\usepackage[utf8]{inputenc}
\usepackage{mathtools}
\usepackage{graphicx}
\graphicspath{{figures/}}
\usepackage{xcolor}
\usepackage{hyperref}
\usepackage{soul}

\newcommand{\gp}{\dot{\gamma}}

\newcommand{\bfn}{\boldsymbol{n}}

\newcommand{\bfx}{\boldsymbol{x}}

\renewcommand{\L}{{\mathcal{L}}}

\newcommand{\txi}{\tilde{\xi}}
\newcommand{\bfxi}{\boldsymbol{\xi}}

\newcommand{\tkappa}{\tilde{\kappa}}

\newcommand{\nablax}{\nabla_{\xi} \, }
\newcommand{\nablaxi}{\nabla_{\xi^i} \, }
\newcommand{\nablatx}{\nabla_{\txi} \, }

\newcommand{\dGamma}{\dot{\Gamma}}

\begin{document}

\title{Curvature-driven transport of thin Bingham fluid layers in airway bifurcations}.

\author{Cyril Karamaoun}
\affiliation{Sorbonne Universit\'e, Laboratoire Jacques-Louis Lions (LJLL), F-75005 Paris, France}
\affiliation{Universit\'e C\^ote d'Azur, LJAD, VADER Center, Nice, France}
\author{Haribalan Kumar}
\affiliation{Auckland Bioengineering Institute, Auckland, New-Zealand}
\author{M\'ed\'eric Argentina}
\affiliation{Universit\'e C\^ote d'Azur, Institut de Physique de Nice, VADER Center, Nice, France}
\author{Didier Clamond}
\affiliation{Universit\'e C\^ote d'Azur, LJAD, VADER Center, Nice, France}
\author{Benjamin Mauroy}
\email{benjamin.mauroy@univ-cotedazur.fr}
\affiliation{Universit\'e C\^ote d'Azur, CNRS, LJAD, VADER Center, Nice, France}

\date{\today}

\begin{abstract}
The mucus on the bronchial wall forms a thin layer of non-Newtonian fluid, protecting the lungs by capturing inhaled pollutants. Due to the corrugation of its interface with air, this layer is subject to surface tension forces that affect its rheology. 
This physical system is analyzed using lubrication theory and 3D simulations. 
We characterize the nonlinear behavior of the mucus and show that surface tension effects can displace overly thick mucus layers in airway bifurcations. 
This movement can disrupt mucociliary clearance and break the homogeneity of the layer thickness.
\end{abstract}

\pacs{}
\maketitle
As one of the central organs of respiration, the main functions of the lung are to bring 
oxygen from ambient air to the blood and to extract carbon dioxide from the body. 
The large exchange surface between the air and the blood, about $75$--$100$ m$^2$ 
\cite{weibel_pathway_1984,west_respiratory_2011}, is connected to the ambient air by a 
space-filling and multi-scale network of airways. 
To perform its functions, the lung relies on several physical processes and on its tree-like geometry. 
The transport of oxygen and carbon dioxide in the lung is the most studied physical process \cite{sapoval_smaller_2002,noel_origin_2022}.
However, other physical processes are involved, and some of them protect the integrity of the organ \cite{weibel_pathway_1984}.

One of these mechanisms relies on a layer of non-Newtonian fluid 
coating the airways walls: bronchial mucus \cite{fahy_airway_2010}. 
The lung is a potential entry point for external contaminants -- dust particles, chemicals, bacteria, viruses -- that are constantly inhaled.
The mucus traps contaminants and is transported towards to the larynx, 
where it is either expelled or swallowed. 
Two main phenomena are responsible for mucus displacement \cite{king_physiology_2006}.
First, ciliated cells located in the bronchial epithelium beat metachronously 
\cite{chateau_why_2019}, with the cilia pushing the mucus toward the trachea.
This phenomenon is called {\it mucociliary clearance} \cite{king_physiology_2006, bustamante-marin_cilia_2017}.
Second, during coughing \cite{macklem_physiology_1974, basser_mechanism_1989} or at high ventilation rates \cite{stephano_wall_2021}, exhaled airflows can apply shear stress to the mucus that is high enough to induce its displacement.
The efficiency of the protection by the mucus depends on the proper functioning of these two phenomena. 
Pathologies such as asthma, chronic obstructive pulmonary disease (COPD), or cystic fibrosis can impair mucociliary clearance, leading to major respiratory symptoms and to infections 
\cite{del_donno_effect_2000, robinson_mucociliary_2002, ito_impairment_2012}. 
The mucociliary clearance and the air-mucus interaction have been explored thoroughly  \cite{basser_mechanism_1989, zahm_role_1991, mauroy_toward_2011, mauroy_toward_2015, karamaoun_new_2018, xu_mathematical_2019, stephano_modeling_2019, gsell_hydrodynamic_2020, stephano_wall_2021}.

Other physical phenomena, such as gravity \cite{romano_effect_2022} and surface tension \cite{erken_capillary_2022}, can affect mucus transport.
The role of surface tension on the air-mucus interface remain not well understood as of today.
The large airway curvatures suggest that the multi-scaled structure of the bronchial tree, together with surface tension and mucus rheology, can affect the transport of mucus. 
However, in most studies, mucus is modeled as a Newtonian fluid \cite{de_gennes_capillarity_2004, manolidis_macroscopic_2016, ogrosky_impact_2021, erken_capillary_2022, romano_effect_2022}.

Surface tension induces (Laplace) pressure jumps, $\Delta p_L$, across curved interfaces ($\Delta p_L 
= 2 \gamma \kappa$, where $\gamma$ is the surface tension coefficient and $\kappa$ the mean curvature 
of the interface). 
The distribution of this pressure can be evaluated in a self-similar tree model of the bronchial tree \cite{weibel_morphometry_1963, weibel_pathway_1984, mauroy_optimal_2004, tawhai_ct-based_2004, sobac_allometric_2019, noel_origin_2022}.
This model is a bifurcating tree with branches as perfect cylinders.
The size of the branches decreases homothetically at each bifurcation by a factor $h =2^{-1/3} \simeq 0.79$.
In this model, the airways are indexed by their generation $i$, representing the number of 
bifurcations from the airway to the root of the tree (trachea).
The radius of the airways in the $i$-th generation is $r_i = h^i \, r_0$, with $r_0$ being the radius of 
the root of the tree.
Assuming that air-mucus interface and the airways have the same curvature, the Laplace pressure in the $i$-th generation is $p_{L,i} = -\gamma / r_i$.
This pressure decreases with the generation index, with curvature effects tending to push the layer toward the deeper parts of the tree. 
The resulting stress in a layer of thickness $\tau$ that coats the bifurcation between generations $i$ and $i$+1 can be evaluated as $\sigma_i \simeq \gamma \frac{h-1}{r_i^2}\frac{\tau}{2}$, see \cite[section I]{karamaoun_curvature-induced-SM_2022}.
This stress is larger in the small bifurcations, as it increases with the generation index.
However, a more detailed analysis is needed to evaluate whether the curvature effects can indeed move the mucus.
Therefore, detailed bifurcation shapes, more realistic mucus rheology, and mucus hydrodynamics must be included in the model.

Mucus is a complex viscoelastic fluid, potentially thixotropic \cite{lai_micro-_2009}.
Its rheological properties depend on the individual, the localization of the mucus in the bronchial tree, and the environmental factors such as air humidity and temperature. 
Mechanical constraints also influence mucus behavior, and one of its core properties is to exhibit a yield stress, $\sigma_y$, below which it behaves like a solid material \cite{lai_micro-_2009, mauroy_toward_2011, mauroy_toward_2015, stephano_modeling_2019, stephano_wall_2021}. 
This characteristic means that the internal shear stresses in the fluid must overcome $\sigma_y$ for the mucus to behave like a fluid with viscosity $\mu$. 
The typical healthy thickness $\tau$ of the mucus layer is on the order of $10 \ \mu$m \cite{karamaoun_new_2019}.
To understand under which conditions surface tension can be large enough to overcome the mucus yield stress, we modeled the mucus as a Bingham fluid.
This approach has already provided valuable insights \cite{mauroy_toward_2011, mauroy_toward_2015}. 
Moreover, it captures the layer's nonlinear dynamics, which cannot be well represented with a Newtonian fluid.

\begin{table}[t!]
\begin{tabular}{lccc}
  \hline
  Quantity & Value (ref value) & Notation & Ref\\
  \hline
  \multicolumn{3}{l}{Lungs' geometrical data}\\
  \hline
  trachea radius & $0.01$ m& $r_0$ & \cite{weibel_pathway_1984}\\
  reduction factor & $\left(1/2\right)^\frac13 \sim 0.79$ & $h$  & \cite{weibel_pathway_1984,mauroy_optimal_2004,tawhai_ct-based_2004}\\
  \hline
  Mucus properties & & &\\
  \hline
  surface tension & $0.03$ Pa\,m & $\gamma$ & \cite{hamed_synthetic_2014}\\
  mucus density & $1000$ kg.m$^3$ (water) & $\rho$ & \cite{ohar_role_2019}\\
  mucus viscosity & $10^{-3}$ - $10$ Pa\,s ($1$) & $\mu$ & \cite{lai_micro-_2009}\\
  mucus yield stress & $10^{-2}$ - $10$ Pa ($0.1$) & $\sigma_y$ & \cite{lai_micro-_2009}\\
  \multicolumn{1}{p{2.7cm}}{mucus layer thickness (healthy)} & $5$ - $30$ $\mu$m ($10$) & $\tau$ & \cite{karamaoun_new_2018}\\
  \multicolumn{1}{p{2.7cm}}{cilia induced mucus velocity} & \multicolumn{1}{p{2.5cm}}{$10$ - $500$ $\mu$m\,s$^{-1}$ ($50$)} & $v_{\rm{cilia}}$ & \cite{fulford_muco-ciliary_1986, manolidis_macroscopic_2016, karamaoun_new_2019}\\
  \hline
\end{tabular}
\caption{Data range of our model parameters and their default values used in this paper, shown between parentheses.}
\label{data}
\end{table}

According to Table \ref{data}, the Reynolds number is small in the mucus layer, 
$\mathrm{Re} \sim \rho v_{\rm{cilia}} \tau / \mu < 0.015$. 
Thus, fluid mechanics can be approximated using the Stokes equations.
Denoting $\boldsymbol{u}$ as the fluid velocity and $p$ as the fluid pressure, the momentum and mass 
conservation equations are:
\begin{equation*}
\begin{array}{ll}
\rho \frac{\partial \boldsymbol{u}}{\partial t} - \nabla \cdot \Sigma + \nabla p = 0
&\text{in the layer}\\
\nabla \cdot \boldsymbol{u} = 0&\text{in the layer}\\
\Sigma \cdot \bfn - p \, \bfn = p_L \, \bfn & \text{at air-fluid interface $\L_t$}\\
\boldsymbol{u} = 0 & \text{at airway wall $W$}\\
\frac{\mathrm{d} \bfx}{\mathrm{d} t} = \left(\boldsymbol{u}(\bfx,t) \cdot \bfn(\bfx,t)\right)\bfn(\bfx,t) & \text{$\bfx \in \L_t$}\\
p_L = - 2 \gamma \kappa(\bfx,t) & \text{$\bfx \in \L_t$}
\end{array}
\label{mom}
\end{equation*}
Here, $\bfn(.,t)$ represents the normal to the air-Bingham fluid interface $\L_t$ at time $t$, and $\kappa$ denotes the mean curvature of this interface. 
The normals are oriented towards the air medium, and the characteristic thickness of the layer along these normals is denoted by $\tau$.
$\Sigma$ is the viscous stress tensor. 
The air-Bingham fluid interface $\L$ is a free surface, and a geometric point $\bfx$ 
on its surface moves with the normal component of the Bingham layer velocity, 
$\left(\boldsymbol{u}(\bfx) \cdot \bfn(\bfx)\right)\bfn(\bfx)$.
Mucociliary clearance could be modeled with a slip boundary condition on the wall $W$. 
However, to isolate the sole effects of surface tension, clearance is not considered in the model, and a non-slip boundary condition is assumed on $W$.

\begin{figure*}[t!]
{\bf A} \includegraphics[height=4.6cm]{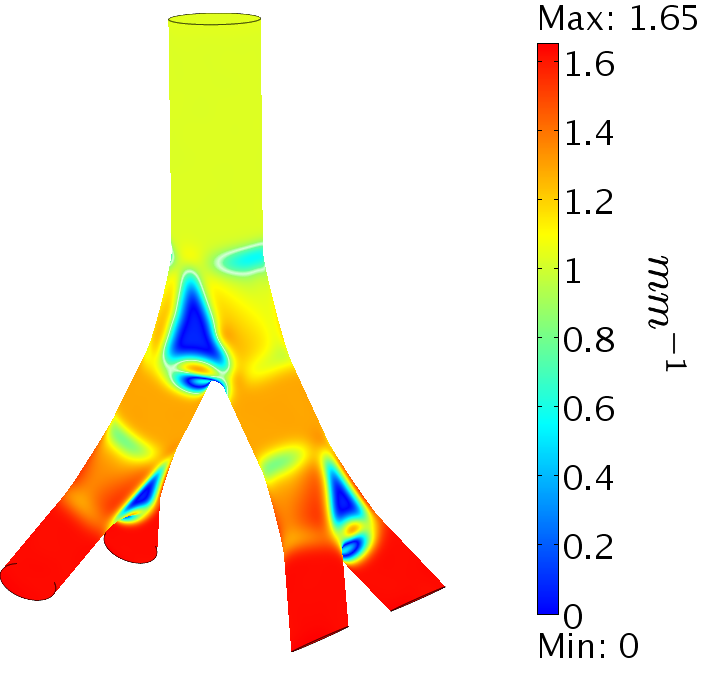}
{\bf B}\hspace{0.2cm}\includegraphics[height=4.6cm]{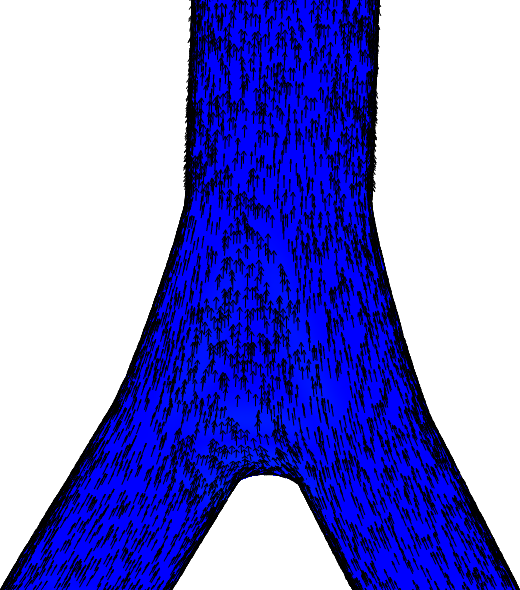}
{\bf C}\hspace{0.2cm}\includegraphics[height=4.6cm]{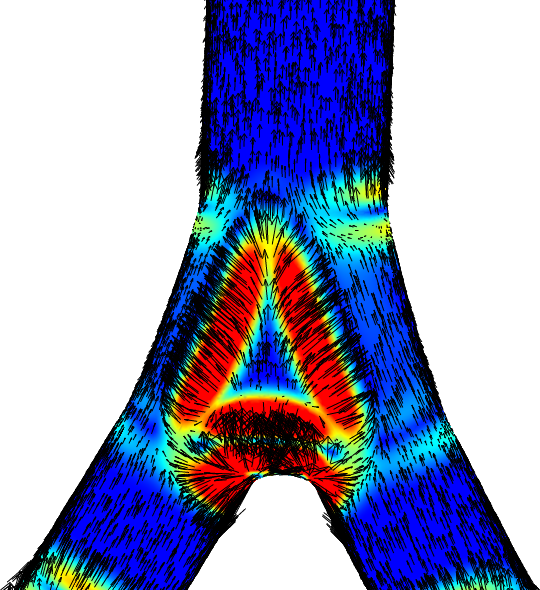}
\includegraphics[height=4.6cm]{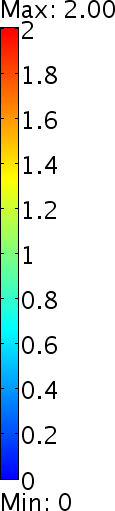}
\caption{{\bf A}: The reference 3D geometry used with the lubrication theory. The geometry is rescaled to cover all the scales of the bronchial tree bifurcations. 
The root branch radius is $1$ mm.
The branches size decreases with a ratio $h = \left( 1/2 \right)^\frac13$ at the bifurcations. 
The branching angle is $60^{\circ}$, and the angle between the two successive branching planes is $90^{\circ}$, in accordance with the mean observed values~\cite{tawhai_ct-based_2004}. 
The colors represent twice the non signed mean curvature field $|\kappa_0|$ (m$^{-1}$). 
{\bf B} ($\tau = 10$ $\mu$m), {\bf C} ($\tau = 75$ $\mu$m): Bingham fluid velocity fields $\boldsymbol{v}_{m,i}$ (arrows) and the ratio $\alpha$ between the amplitudes of the velocity induced by surface tension effects and the velocity induced by the idealized mucociliary clearance, i.e. $\alpha = \| \boldsymbol{v}_{\textrm{st,}i} \| /\| \boldsymbol{v}_{\textrm{cilia}} \|$ (colors). 
Here, the mother branch has a radius of $1$ mm and the daughter branches a radius of $\left(1/2\right)^{1/3} \simeq 0.79$ mm. 
At physiological thickness (B), only the idealized mucociliary clearance drives the motion of the mucus. 
For non healthy thickness (C), the idealized mucociliary clearance is strongly altered by surface tension effects.
}
\label{veloTreeMucus}
\end{figure*}

The Bingham fluid constitutive equations are
\begin{equation*}
\left\{
\begin{array}{ll}
\Sigma = \left( \mu + \frac{\sigma_y}{\gp} \right) \dGamma & \text{for } \sigma > \sigma_y\\
\dGamma = 0  & \text{for } \sigma \leq \sigma_y
\end{array}
\right.
\label{bing}
\end{equation*}
with $\dGamma = \frac12 \left( \nabla u + \left(\nabla u\right)^{\rm{t}} \right)$ the rate of strain tensor and with $\sigma = \sqrt{ \frac12 \Sigma{:}\Sigma }$ 
and $\gp = \sqrt{ \frac12 \dGamma{:}\dGamma }$ the second invariants of, respectively, the stress tensor and the rate of strain tensor.

Given that the mucus layer is generally thin compared to the mean curvature radius of the bronchial wall, planar lubrication theory applies \cite{panton_incompressible_2013, balmforth_consistent_1999} (see \cite[section II]{karamaoun_curvature-induced-SM_2022}). 
Our results show that the viscous stresses tangential to the wall dominate inside the fluid.
These stresses are proportional to the interface curvature gradient $\nablax \kappa$, computed in the curvilinear coordinate system $\bfxi$. 
The normal stress is dominated by the Laplace pressure.
Our theory uncovers a characteristic Bingham-like number $B = \frac{\sigma_y r^2}{2 \gamma \tau}$, where $r$ is the characteristic radius of curvature of the interface, typically the radius of the airway.
The fluid layer remains liquid if $B < \| \nablatx \tkappa \|$, where $\nablatx \tkappa = r^2 \times \nablax \kappa$ (quantities with tildes are normalised by $r$).

A proportion $e = \max \left( 0, 1 - {B}/{\| \nablatx \tkappa \|}\right)$ of the layer near the wall is liquid \cite{mauroy_toward_2011, mauroy_toward_2015}.
Above this liquid layer, the Bingham fluid is solid.
The liquid layer drags the solid one. 
The surface tension homogenizes the layer thickness, thus we assume hereafter that $\tau$ is constant and that the layer curvature equals that of the airways (this analysis assumes the airways are perfectly smooth).
Thus, we do not account for large fluid accumulations, clots or plugs \cite{grotberg_respiratory_2001, mauroy_toward_2011, mauroy_toward_2015, stephano_modeling_2019}.
Under these conditions, for the $i$-th generation and to the leading order, the velocity field $\boldsymbol{v}_{\textrm{st,} i}$ (averaged over the layer thickness) is
\vspace{-0.05cm}
\begin{equation}
    \begin{array}{l}
    \begin{aligned}
    &\boldsymbol{v}_{\textrm{st,} i} = \ - f(e_i) \ \frac{\gamma \tau^2}{\mu r_i^2}  \ \nablatx \tkappa, \quad f(e) = e^2 \left( 1 - \frac{e}{3} \right),
 \\
    &e_i = \max \left( 0, 1-\frac{B_i}{\| \nablatx \tkappa \|}\right).
        \end{aligned}
    \end{array}
    \label{meanVelMain}
\end{equation}
The vector $-\frac{\gamma \tau^2}{\mu r_i^2} \nablatx \tkappa$ represents the velocity in the case of a Newtonian fluid.
Our results show the presence of a prefactor $f(e_i)$ in the case of a non-Newtonian fluid. 
This dimensionless prefactor fully characterizes the non-Newtonian behavior and depends nonlinearly on the Binghman number $B_i$ in the $i$-th generation.

Due to the scaling law between generations, the mean curvature gradients in the $i$-th generaton are $ \nablaxi \kappa_i = h^{-2i} \ r_0^{-2} \ \nabla_{\txi} \tkappa$. 
This leads to a scaling law for the fluid velocity:
\begin{equation}
   \boldsymbol{v}_{\textrm{st,} i} =\left(\frac1{ h^{2}}\right)^i f(e_i) \ \frac{\gamma \tau^2}{\mu {r_0}^2 } \nabla_{\txi} \tkappa.
\end{equation}
In addition, the wall mean curvature $\tkappa$ is obtained from an idealized 3D tree 
geometry, see Fig. \ref{veloTreeMucus}A and \cite[section IV]{karamaoun_curvature-induced-SM_2022}.
The fluid velocity field in the 3D idealized tree is obtained from the theoretical formula (\ref{meanVelMain}). 
Finally, the velocity induced by mucociliary clearance is also evaluated in the 3D idealized tree.
It is modeled as the gradient of a Laplacian field on the bifurcation walls, see \cite[section V]{karamaoun_curvature-induced-SM_2022}.
The resulting idealized clearance is tangential to the bifurcation walls and directed toward the larger airways \cite{manolidis_macroscopic_2016}.
Its amplitude is set to $50$ $\mu$m.s$^{-1}$ \cite{karamaoun_new_2019}.

\begin{figure}[t!]
\centering 
\includegraphics[width=8cm]{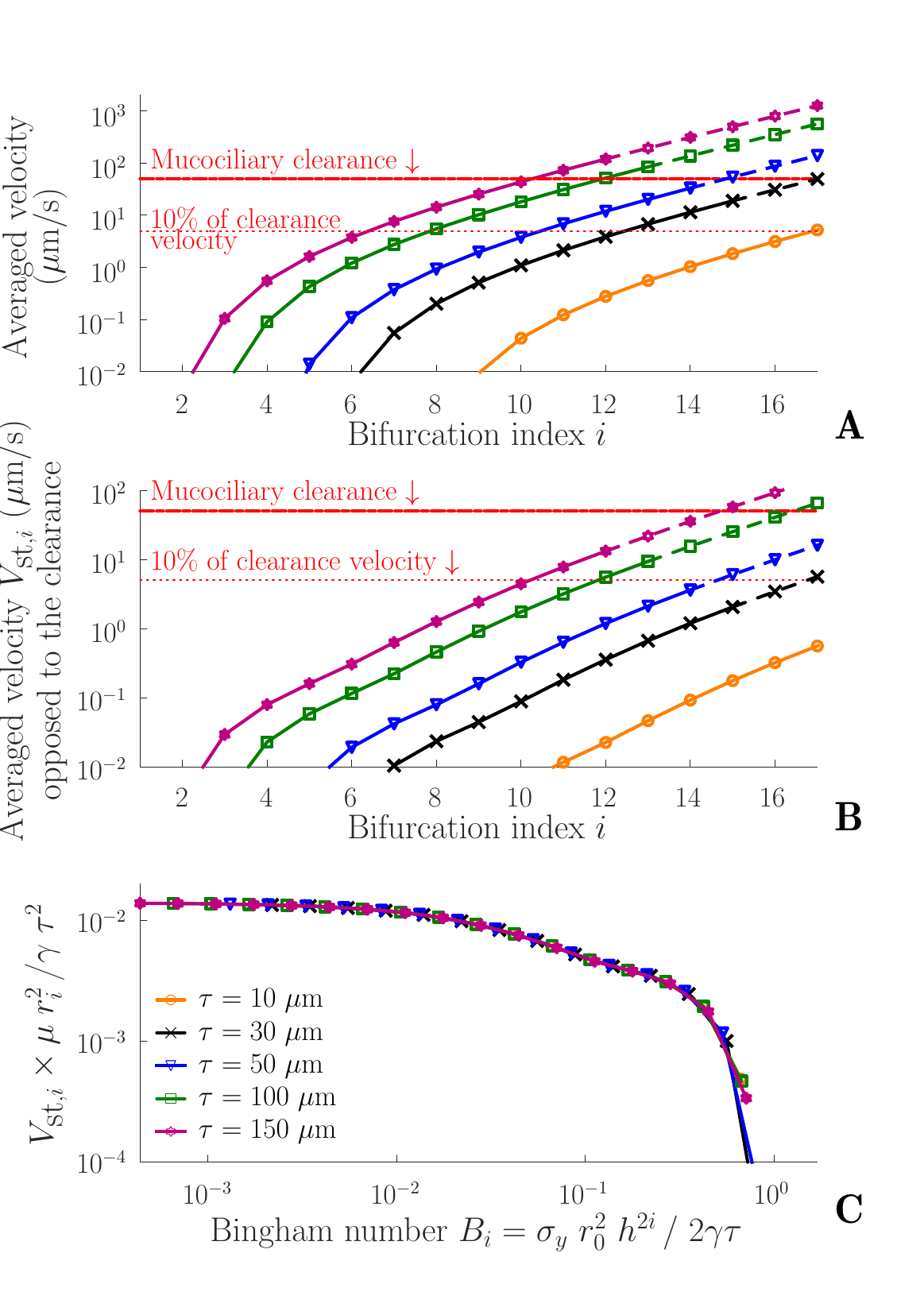}
\caption{Velocity patterns with parameters values $\sigma_y = 0.1$ Pa, $\mu = 1$ Pa\,s and a clearance amplitude is $50 \ \mu$m\,s$^{-1}$.
The dashed parts of the curves correspond to generations where the model hypothesis $\tau / r_i \ll 1$ loses its validity (i.e. $\tau / r_i > 10 \%$).
Our results show that curvature effects can disrupt mucociliary clearance for pathological thicknesses of the layer ($\tau \geqslant 50 \ \mu$m), particularly in the medium and small airways.
{\bf A:} Bingham layer velocity amplitude (curvature effects only) averaged over the bifurcation and layer thickness.
{\bf B:} Component of the Bingham layer velocity (curvature effects only) in the direction of the idealized clearance and averaged over the bifurcation and layer thickness.
The average velocity induced by curvature effects is always opposed to clearance.
{\bf C:} Averaged velocity opposed to clearance, normalized by the velocity of a Newtonian fluid in the same configuration.
The normalized velocity depends only on the Bingham number $B_i$.} 
\label{vstPlot}
\end{figure}

Many common lung pathologies induce a thickening of the layer \cite{williams_airway_2006}. 
Therefore, in addition to healthy mucus layer thickness, other thicknesses compatible with mucus pathophysiology were tested.
Typical velocity outputs of this study are presented in Fig. \ref{veloTreeMucus} (spatial distribution) and Fig. \ref{vstPlot}A (averaged amplitude).
Our results show that after a threshold generation, the curvature gradients are able to overcome the yield stress of the Bingham fluid.
This implies that, from this generation onward, the fluid exhibits liquid behavior and can be displaced. 
For healthy mucus ($\tau = 10 \ \mu$m), this threshold is around the $9^{\text{th}}$ generation, see Fig. \ref{vstPlot}A.
In this case, the velocity magnitude is negligible compared to that induced by mucociliary clearance.
For thicker layers, the threshold generation decreases and the velocity increases, eventually becoming larger than the magnitude of the velocity induced by idealized clearance, see \ref{vstPlot}A.
Thus, the thicker the mucus, the more the idealized clearance is impaired, see Fig. \ref{veloTreeMucus}.
Eventually, the effects of the clearance disappear, and the velocity fields is driven solely by curvatures (Fig. \ref{veloTreeMucus}C). 

The disruption of the clearance depends on the layer thickness and the position in the bifurcation.
The convergence (or divergence) of the predicted velocity field indicates possible mucus accumulation (or depletion) at different spots in the bifurcation, see Fig. \ref{veloTreeMucus}C.
A local accumulation increases the risk of bronchial obstruction. 
Conversely, a local depletion reduces the protection of the epithelium, making it more susceptible to external contaminants and physicochemical stresses \cite{karamaoun_is_2022}. 
Furthermore, our model suggests that local mucus accumulation is likely to develop first in the parts of the airways closest to the bifurcation zone, see Fig. \ref{veloTreeMucus}C.
On the other hand, the carina of the bifurcation, i.e., the meeting point of the two small airways, is more susceptible to mucus depletion since it has relatively low curvature (Fig. \ref{veloTreeMucus}A). 
Notably, inhaled particules are more likely to deposit at the carina \cite{balashazy_local_2003, zierenberg_asymptotic_2013}, and mucus overproduction might counterintuitively increase the risk of epithelial damage near the carina \cite{balashazy_local_2003}. 

To quantify how the curvature gradient opposes the mucociliary clearance in a bifurcation, we project the velocity field $v_{\textrm{st},i}$ (induced by surface tension) onto the local direction of the idealized clearance.
Then, we average this velocity component over the layer thickness and the bifurcation to define $V_{\textrm{st},i}$, see \cite[section III]{karamaoun_curvature-induced-SM_2022}.
The results are shown in Fig. \ref{vstPlot}B and C.
In all the cases, the averaged Bingham fluid velocity opposes that induced by the idealized clearance.
Moreover, the non-Newtonian effects are driven only by the Bingham number $B_i =  \frac{\sigma_y r_i^2}{2 \gamma \tau}$, as shown in Fig. \ref{vstPlot}C.
For $B_i < 0.1$, the fluid behaves as a liquid throughout the entire tree and most strongly opposes clearance.
For $B_i > 0.5$, the velocities drop drastically, and the clearance is no longer disrupted; in this case, the Bingham fluid is no longer fully liquid and eventually behaves as a solid.

These results agree with observations, as an increased thickness of the mucus layer in pathological conditions has been associated with a disturbance of clearance \cite{del_donno_effect_2000, robinson_mucociliary_2002, ito_impairment_2012}. 
Moreover, our analysis highlights the importance of the mechanisms that control mucus thickness, particularly in the bifurcations \cite{karamaoun_new_2019}.

% HERE %

In the absence of clearance, the fluid in the bifurcation is globally driven by capillarity toward the small airways, as shown in Fig. \ref{vstPlot}B.
This movement opposes that induced by clearance.
Curvature gradients vanish in cylindrical airways, and further mucus displacement along the small airways can only result from a non constant thickness of the layer.
Surface tension effects tend to homogenize the layer thickness until the shear stress falls below the yield stress, resulting in local accumulation of mucus.
Thus, our analysis suggests that clearance can also compensate for curvature effects on the mucus in bifurcations. 
Moreover, the curvature gradients trigger the displacement of the layer mainly along the wall, indicating their contribution to maintaining a layer with constant thickness. 
However, when the layer is too thick, the curvature gradients become too strong for clearance to counteract these effects.

Our analysis reveals curvature effects on a Bingham fluid layer coating the airways walls of a bifurcation.
Our results suggest that pathological thickening of the bronchial mucus layer can counteract and potentially disrupt mucociliary clearance.
However, the curvature effects on the mucus remain intricate, and the model does not account for the full complexity of the bronchial mucus layer.

The rheology of mucus remains incompletely understood \cite{jory_mucus_2022}, with properties varying widely between individuals and influencd by environmental factors \cite{lai_micro-_2009, jory_mucus_2022}.
Therefore, our results are primarily qualitative and not exhaustive. 
However, employing a Bingham model to represent mucus allows to capture its key properties, such as viscosity and yield stress \cite{lai_micro-_2009, jory_mucus_2022}, and our predictions align with lung pathophysiology \cite{del_donno_effect_2000, robinson_mucociliary_2002, ito_impairment_2012}.

The thickness of mucus likely varies depending on location within the bronchial tree \cite{karamaoun_new_2019}.
Moreover, bronchi are not perfect cylinders, and their wall curvature is not constant. 
These phenomena, which are not considered in our model, can influence the layer thickness. 

Our study demonstrates that curvature effects can displace thick layers of Bingham fluid in airways bifurcations.
In the context of lung pathologies, our findings suggest that curvature effects likely play a significant role in disrupting mucociliary clearance when mucus accumulation is present.
This work paves the way to a deeper understanding of bronchial mucus dynamics in pathological lungs.
Future research should incorporate more realistic models of the bronchial tree and air--mucus interface, as well as consider other biophysical phenomena such as gravity \cite{romano_effect_2022}, to assess their respective impacts.
Moreover, our work highlights the possibility of ``hidden'' physical processes being triggered by certain pathologies, significantly affecting organs function.

%%%%%%%%%%%%%%%%%%%%%%%%%%%%%%%%%%%%%%%%%%%%%%%%%%%%%%%%%%%%%%%%%%
\begin{acknowledgments}
This work has been supported by the Mission pour l'interdisciplinarité du CNRS, the Agence Nationale de la Recherche (the VirtualChest project, ANR-16-CE19-0014; the IDEX UCA JEDI, ANR-15-IDEX-01), the Académie des Systèmes Complexes de l'Université Côte d'Azur and the association Vaincre La Mucoviscidose (RF20190502489).
\end{acknowledgments}
\newpage
\bibliographystyle{abbrv}   
%\bibliographystyle{unsrtnat}
%\bibliography{bibli_2023}

\end{document}

\title{Supplemental Materials for\\ Curvature-driven transport of thin Bingham fluid layers in airway bifurcations.}.

\author{Cyril Karamaoun}
\affiliation{Université Côte d'Azur, LJAD, VADER Center, Nice, France}
\author{Haribalan Kumar}
\affiliation{Auckland Bioengineering Institute, Auckland, New-Zealand}
\author{Médéric Argentina}
\affiliation{Université Côte d'Azur, Institut de Physique de Nice, VADER Center, Nice, France}
\author{Didier Clamond}
\affiliation{Université Côte d'Azur, LJAD, VADER Center, Nice, France}
\author{Benjamin Mauroy}
\email{benjamin.mauroy@unice.fr}
\affiliation{Université Côte d'Azur, CNRS, LJAD, VADER Center, Nice, France}

\date{\today}

\maketitle

\tableofcontents

\newpage

\section{Analysis in a fractal tree}
\label{fractal}

\begin{figure}[t!h]
\centering 
\includegraphics[width=5cm]{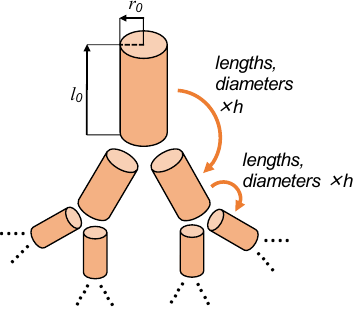}
\caption{Fractal model of the bronchial tree used in the qualitative analysis. 
The size of the branches is decreasing at each bifurcation by a factor $h = (1/2)^{\frac13}$. 
The radius and length of the tree root (trachea) are respectively $r_0$ and $l_0$.} 
\label{fractalFig}
\end{figure}

We seek a qualitative estimation of the pressures induced by surface tension in the generations of the airway tree. 
The geometry of the bronchial tree is approximated as a cascade of bifurcating cylindrical airways~\cite{mauroy_interplay_2003, mauroy_optimal_2004}, as shown in Fig. \ref{fractalFig}.
The airways are numbered by using a generation index $i$, which represents the number of bifurcations from the root of the tree, i.e., the trachea, to the considered airway. 
We assume that the dimensions of the airways between two consecutive generations are related by a homothetic factor $h$, independent of the generation index. 
The theoretical value $h = (1/2)^{1/3} \simeq 0.79$ has been found to adequately represent the geometry of the mammalian lung \mbox{\cite{weibel_morphometry_1963, weibel_pathway_1984, mauroy_optimal_2004, tawhai_ct-based_2004, sobac_allometric_2019, noel_origin_2022}}.

We assume that the mucus layer in this geometry has a negligible thickness relative to the airways radii \cite{weibel_pathway_1984, karamaoun_new_2019}. 
The principal curvatures of the air--Bingham fluid interface in generation $i$ can then be approximated by the principal curvatures of the cylindrical airway: $1/r_i$ in the radial direction and $0$ in the axial direction.
These curvatures induce a Laplace pressure drop $p_{L,i}$ between the air and the Bingham fluid:
$$
p_{L,i} = - \frac{\gamma}{r_i}
$$
Since the airways are considered perfect cylinders, the radius within a single bronchus does not vary.
Thus, there is no gradient of Laplace pressure, and the Bingham fluid is motionless. 
However, the radii vary between the airways, as the distal (deep) bronchi are smaller than the proximal (upper) ones. 
Because of this change in curvature, the amplitude of the pressure drop increases with the generation index. 
This implies that a pressure gradient exists between two successive generations, which are connected through bifurcations.

Between two successive generations $i$ and $i+1$, the radii $r_i$ and $r_{i+1}$ are related as $r_{i+1} = h \times r_i$. 
Assuming that the length of the bifurcation $\Delta x$ is of the order of magnitude as the airway radius, the curvature radius gradient between two successive generations can be approximated by $\frac{\Delta r_i}{\Delta x} \simeq \frac{h \times r_i - r_i}{r_i} = (h-1) < 0$. 
Hence, as in~\cite{mauroy_toward_2011}, the mean shear stress applied to the layer by the pressure drop between two successive generations $i$ and $i+1$ can be qualitatively evaluated as
$$
\sigma \simeq \frac{\Delta p_{L,i}}{\Delta x} \frac{\tau}{2} \simeq \gamma \frac{h-1}{r_0^2 h^{2i}}\frac{\tau}{2} < 0
$$
If this stress overcomes the yield stress $\sigma_y$, the Bingham fluid flows. 
Because $\Sigma_{rx}$ is negative, the Bingham fluid should flow toward the distal regions of the tree, opposite to the direction of mucociliary clearance.

The previous analysis suggests that the Laplace pressure gradients should be stronger in the distal bifurcations than in the proximal bifurcations. 
However, a more refined analysis is needed to obtain accurate estimations of those gradients and to determine if they can induce shear stresses high enough to overcome the yield stress of the Bingham fluid.

%%%%%%%%%%%%%%%%%%%%%%%%%%%%%%%%%%%%%%%%%%%%%%%%%%%%%%%%%%%%%%%%%%%%%%

\section{Lubrication theory, Bingham case}
\label{lubTheo}

\subsection{Local coordinates system}
\label{coords}

\noindent{\bf Coordinates change.}\\
To derive the main components of the velocity in the Bingham layer, we will use a lubrication technique based on the hypothesis that the thickness of the layer $\tau$ is much smaller than the characteristic length of the domain~\cite{balmforth_consistent_1999}.
This characteristic length is estimated using the characteristic curvature radius $R$ of the surface on which the layer spreads. 
Typically, this characteristic curvature radius corresponds to the radius of the airway considered.

\begin{figure}[h!]
\centering 
\includegraphics[width=9cm]{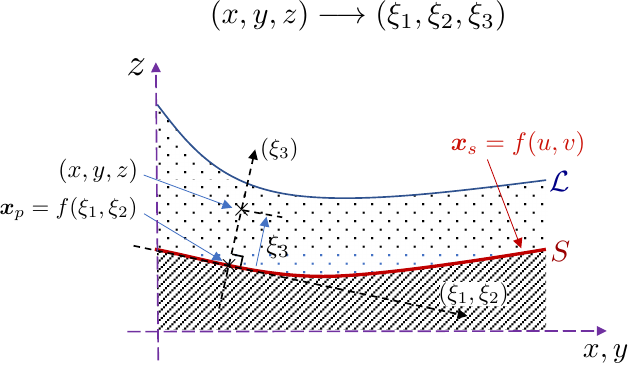}
\caption{Transformation from global coordinates $\bfx = (x,y,z)$ to local coordinates $\bfxi = (\xi_1,\xi_2,\xi_3)$.
The plane $(\xi_1 \xi_2)$ is tangent to the wall $S$ defined by the surface $\bfx_S = f(u,v)$, where $(u,v)$ is a set of curvilinear coordinates.
The direction $(\xi_3)$ is normal to $S$.
The surface $\L$ represents the air--Bingham fluid interface.} 
\label{flatCoord}
\end{figure}

The first step is to use a local coordinates system.
We will denote $(x,y,z)$ as the physical coordinates and $(\xi_1,\xi_2,\xi_3)$ as the local coordinates system, as schematized in Fig. \ref{flatCoord}. 

We will consider a thin layer that stands on a substrate.
The surface $S$ of the substrate is represented locally by a parametric representation $\bfx_S = f(u,v)$, where $(u,v) \in \Omega$ is a curvilinear parameterization of the surface and $\Omega$ a subset of $\mathbb{R}^2$. 
In this case, we can project a point $\bfx = (x,y,z)$ in the layer onto the substrate surface, see Fig. \ref{flatCoord}. 
The resulting projection point on the surface is denoted $\bfx_p = f(\xi_1,\xi_2)$.
Then,
$$
\bfx = f(\xi_1,\xi_2) + \xi_3 \ \bfn_S
$$
with
$$
\bfn_S = \left(\frac{\partial f}{\partial u} \wedge \frac{\partial f}{\partial v} \right) \left/ \left| \left|\frac{\partial f}{\partial u} \wedge \frac{\partial f}{\partial v}\right| \right| \right.
$$
The new coordinates system is then determined by the triplet $\bfxi = (\xi_1,\xi_2,\xi_3)$.

We will now use a dimensionless formulation of the equations in order to characterize the dominant velocity of the mucus when $\epsilon = \tau/R$ is small relatively to $1$.
The ratio $\epsilon$ represents the thickness of the mucus layer relatively to the curvature of the airways.

We define dimensionless coordinates associated to the triplet $\bfxi = (\xi_1, \xi_2, \xi_3)$ using the new triplet $\bftxi = (\txi_1,\txi_2,\txi_3)$ with $\xi_1 = R \txi_1$,  $\xi_2 = R \txi_2$ and $\xi_3 = \tau \txi_3$.
In the following, the notation with a tilde over a letter indicates a dimensionless quantity.

\begin{itemize}
\item Since $f$ defines the airway wall, we assume that the characteristic size of $f$ is also $R$ and we define $\tf$ as $f(u,v) = R \tf(\tu,\tv)$ with $\tu = u/R$ and $\tv = v/R$.

\item The Laplace pressure is normalized based on $P = \gamma/R$ and $\tp_L = p_L/P$. 
Moreover, for newtonian fluids, the velocity is proportional to $\frac{\gamma}{\mu} \epsilon^2$, hence we choose the scaling $U = \frac{\gamma}{\mu} \epsilon^2$ for the velocity components in $\xi_1$ and $\xi_2$ directions. 
In the $\xi_3$ direction, the scaling of the velocity is $W = U \tau / R = U \epsilon$.

\item Stresses are rescaled with $\sigma^{*3} = \mu \frac{U}{\tau} \tsigma^{*3}$ for $* = 1$ or $2$, and the other components are rescaled with $\sigma_{*} = \mu \frac{U}{R} \tsigma{*}$ for $*=11$, $22$, $33$ and $12$. 
We denote $\Sigma_{i,j}$ the rescaled value of $\sigma^{ij}$.

\item Strains are rescaled accordingly, i.e. as $\sigma / \mu$.

\item We assume that all dimensionless variables can decompose into a series relatively to $\epsilon$, i.e. for a variable $*$, its decomposition writes $* = *_0 \ \epsilon^0 + *_1 \ \epsilon + *_2 \ \epsilon^2 + ...$.\\
\end{itemize}

\subsection{Metric}
\label{metricTensor}

We define the matrix $C = (c_{i,j})_{i,j} = (\bfb_1, \bfb_2, \bfb_3)$ with 
$$
\begin{array}{l}
\bfb_1 = \frac{\partial \bfx}{\partial \xi_1} = \frac{\partial f}{\partial u} + \xi_3 \frac{\partial \bfn_S}{\partial u}
= \frac{\partial \tf}{\partial \tu} + \epsilon \txi_3 \frac{\partial \bfn_S}{\partial \tu}\\
\bfb_2 = \frac{\partial \bfx}{\partial \xi_2} = \frac{\partial f}{\partial v} + \xi_3 \frac{\partial \bfn_S}{\partial v}
= \frac{\partial \tf}{\partial \tv} + \epsilon \txi_3 \frac{\partial \bfn_S}{\partial \tv}\\ 
\bfb_3 = \frac{\partial \bfx}{\partial \xi_3} = \bfn_S
\end{array}
$$
and the matrix $\bC = (\bc_{i,j})_{i,j} = C^{-1} = (\bfb^1, \bfb^2, \bfb^3)$.
We denote $C^0$ and $\tC^0$ the first term of the development of $C$ and $\tC$ in $\epsilon$.

The associated metric tensor is defined with $g_{(i,j)} = (\bfb_i . \bfb_j)_{i,j}$ and its inverse with $g^{(i,j)} = (\bfb^i . \bfb^j)_{i,j}$.
The metric tensors can be rewritten in dimensionless coordinates:
$$
g_{(i,j)}(\xi_1, \xi_2, \xi_3) = \tg_{(i,j)}(\txi_1, \txi_2, \txi_3) = \left(
\begin{array}{ccc}
\left\|\frac{\partial \tf}{\partial \tu}\right\|^2 & \frac{\partial \tf}{\partial \tu}\cdot\frac{\partial \tf}{\partial \tv} & 0 \\
\frac{\partial \tf}{\partial \tu}\cdot\frac{\partial \tf}{\partial \tv} & \left\|\frac{\partial \tf}{\partial \tv}\right\|^2 & 0 \\
0 & 0 & 1
\end{array}
\right)
+
O(\epsilon)
$$
and
$$
g^{(i,j)}(\xi_1, \xi_2, \xi_3) = \tg^{(i,j)}(\txi_1, \txi_2, \txi_3) = \frac1{\td(\tu,\tv)} \left(
\begin{array}{ccc}
\left\|\frac{\partial \tf}{\partial \tu}\right\|^2 & -\frac{\partial \tf}{\partial \tu}\cdot\frac{\partial \tf}{\partial \tv} & 0 \\
-\frac{\partial \tf}{\partial \tu}\cdot\frac{\partial \tf}{\partial \tv} & \left\|\frac{\partial \tf}{\partial \tv}\right\|^2 & 0 \\
0 & 0 & \td(\tu,\tv)
\end{array}
\right)
+
O(\epsilon)
$$
with $\td(\tu, \tv)$ the determinant of the first term $\tC^0$ of the development in series of the matrix $\tC$, $\td(\tu,\tv) = \left\|\frac{\partial \tf}{\partial \tu}\right\|^2 \left\|\frac{\partial \tf}{\partial \tv}\right\|^2 - \left(\frac{\partial \tf}{\partial \tu}.\frac{\partial \tf}{\partial \tv}\right)^2$.

For the sake of notation simplicity, we assume in the following that $g^{ij}$, $g_{ij}$, $\tg^{ij}$ and $\tg_{ij}$ refer to the coeffcient associated to $\epsilon^0$ in their decomposition in powers of $\epsilon$.

\subsection{Christoffel symbols of the second kind}

The Christoffel symbols of the second kind $\Gamma_{ik}^j$ allow to compute the derivatives in the local coordinates system $(\xi_1, \xi_2, \xi_3)$.
The symbols are written, using Einstein notation,
$$
\Gamma_{ij}^k = 
\frac12 g^{ip} \left( \frac{\partial g_{pj}}{\partial \xi_k} + \frac{\partial g_{pk}}{\partial \xi_j} - \frac{\partial g_{jk}}{\partial \xi_p}\right) 
$$
We define normalized versions of the Christoffel symbols according to the dimensionless coordinates defined above:
$$
\tCh_{i,j}^k = R \ \Ch_{i,j}^k
$$
The terms in $\epsilon^0$ in the developments in series of the metric tensors $\tg^{(i,j)}$ and $\tg_{(i,j)}$ do not depend on $\txi_3$ and have several null terms in their expression
Moreover, we have $g_{3,3} = 1$. 
With these properties, we can get information on the developments in series of the Christoffel symbols relatively to $\epsilon$:
\begin{equation}
\begin{array}{ll}
\tCh_{i,j}^k = O(\epsilon) & \text{if at least one of $i$, $j$ or $k$ is equal to $3$}\\
\tCh_{i,j}^k = O(1) & \text{otherwise}
\end{array}
\label{chrisnorm}
\end{equation}

\subsection{Equations of the mucus dynamics}

We assume that the layer stands on the airway wall $S$. 
The air--fluid interface is denoted $\L$.
The mucus dynamics equations in the coordinates frame $(x,y,z)$ are
\begin{equation}
\left\{
\begin{array}{ll}
\rho \frac{\partial \boldsymbol{u}}{\partial t} - \nabla\cdot\Sigma = \nabla p&\text{in the layer}\\
\nabla\cdot\boldsymbol{u} = 0&\text{in the layer}\\
\Sigma \cdot \bfn - p \, \bfn = p_L \, \bfn & \text{at the air--fluid interface $\L$}\\
\boldsymbol{u} = 0 & \text{on the airway wall $S$}\\
\frac{\partial \bfx}{\partial t} = \left(\boldsymbol{u}(\bfx)\cdot\bfn\right)\bfn & \text{at the air--fluid interface $\L$, for $\bfx \in \L$}\\
p_L = - 2 \gamma \kappa(\bfx,t) & \text{at the air--fluid interface $\L$}
\end{array}
\right.
\label{momApp}
\end{equation}
We decompose these equations in the coordinates system $(\bfb_1, \bfb_2, \bfb_3)$. 
The coordinates of the velocity $\bfu$ in that frame is $(u^1, u^2, u^3)$.
The covariant differentiation of a quantity $*$ relatively to the coordinate on the component $\bfb_j$ is denoted $*_{,j}$, and 
$$
\begin{array}{l}
p_{,j} = g^{ji} \frac{\partial p}{\partial \xi_i}\\
\Div_{co} \ \Sigma = \sum_{j=1}^3 \left(\frac{\partial \sigma^{ij}}{\partial \xi_i} + \Gamma_{il}^i \sigma^{lj} + \Gamma_{il}^j \sigma^{il} \right) \bfb_j\\
\frac{\partial \bfu}{\partial t} = \sum_{j=1}^3 \frac{\partial u^j}{\partial t} \bfb_j
\end{array}
$$
For any $j \in \{ 1,2,3\}$, the component of the equation on $\bfb_j$ is
$$
\rho \frac{\partial u^j}{\partial t} - \left(\frac{\partial \sigma^{ij}}{\partial \xi_i} + \Gamma_{il}^i \sigma^{lj} + \Gamma_{il}^j \sigma^{il} \right) + g^{ji} \frac{\partial p}{\partial \xi_i} = 0
$$
Finally, we can write the stress tensor as
\begin{equation}
\Sigma = \frac{\mu U}{R} \tSigma \ \text{ with } \ \tSigma = \left(
\begin{array}{ccc}
\tsigma^{11} & \tsigma^{12} & \frac{1}{\epsilon} \tsigma^{13}\\
\tsigma^{12} & \tsigma^{22} & \frac{1}{\epsilon} \tsigma^{23}\\
\frac{1}{\epsilon} \tsigma^{13} & \frac{1}{\epsilon} \tsigma^{23} & \tsigma^{33}\\
\end{array}
\right)
\label{stress}
\end{equation}

\subsection{Air--mucus interface $\L$}

The air--mucus interface $\L$ is defined in the coordinates system $(\xi_1, \xi_2, \xi_3)$ as
$$
\L = \left\{ X_\L = g(u,v) = f(u,v) + \tau \left( 1 + \eta(u,v,t) \right) \bfn_S(u,v) \left| (u,v) \in \Omega \right. \right\}
$$
where $\eta$ is a function from $\Omega$ such that $\eta = o\left( \frac1{\epsilon} \right)$, i.e. $\epsilon \times \eta \underset{\epsilon \rightarrow 0}{\longrightarrow} 0$.
Hence, the order of magnitude of $\eta$ is at most $1$.
Moreover, we assume that the layer cannot be of negative thickness and $\eta \geq -1$.

The normal to the air--mucus interface can then be defined as
$$
\bfn_\L = \left(\frac{\partial g}{\partial u} \wedge \frac{\partial g}{\partial v} \right) \left/ \left| \left|\frac{\partial g}{\partial u} \wedge \frac{\partial g}{\partial v}\right| \right| \right.
$$
Using the dimensionless system of coordinates, we normalize $g$ with $\tg(\txi_1, \txi_2, \txi_3) = g(\xi_1, \xi_2, \xi_3)/R$ and we denote $\tbfn_*(\txi_1, \txi_2, \txi_3) = \bfn_*(\xi_1, \xi_2, \xi_3)$ with $* = \L$ for the normal to the surface $\L$ or $* = S$ for the normal to the surface $S$.
We can relate the normals to the surfaces $S$ and $\L$ with
\begin{equation}
\tbfn_\L = \tbfn_S + \epsilon \underbrace{(1+\teta) \frac{\frac{\partial \tf}{\partial \tu} \wedge \frac{\partial \bfn_S}{\partial \tv} - \frac{\partial \tf}{\partial \tv} \wedge \frac{\partial \bfn_S}{\partial \tu}}{\left| \left|\frac{\partial f}{\partial u} \wedge \frac{\partial f}{\partial v}\right| \right|}}_{\tbfm_\L} + \, O(\epsilon^2)
\label{nL}
\end{equation}
with $\teta(\txi_1, \txi_2, \tt) = \eta(\xi_1, \xi_2, t)$.

The air--mucus interface is a surface of equation $\xi_3 = \tau ( 1 + \eta(\xi_1, \xi_2) )$ or, in dimensionless coordinates, $\txi_3 = 1 + \teta(\txi_1, \txi_2)$.\\

\subsection{Boundary conditions at the air--mucus interface $\L$}

Based on equation (\ref{nL}), the boundary condition $\Sigma \cdot \bfn_\L - p \, \bfn_\L = - p_L \, \bfn_\L$ at the air--mucus interface becomes in the dimensionless formulation 
$$
\tSigma \cdot \tbfn_\L - \frac{1}{\epsilon^2} \tp \, \tbfn_\L = - \frac{1}{\epsilon^2} \tp_L \, \tbfn_\L
$$
Then, based on the expression of $\tSigma$ in equation (\ref{stress}) and of the normal at the air--mucus interface in equation (\ref{nL}), $\tbfn_\L = \tbfn_S + \epsilon \ \tbfm_\L + O(\epsilon^2)$, we can derive the following relationships,
\begin{itemize}
\item At the order $\frac1{\epsilon^2}$: $\tp^0(\xi_1, \xi_2,1+\teta(\xi_1,\xi_2, \tt)) = \tp_L(\xi_1, \xi_2)$.
\item At the order $\frac1{\epsilon}$: 
the boundary condition at $\txi_3 = 1 + \eta(\txi_1, \txi_2, \tt)$ is, at the order $1/\epsilon$: 
$$
\left(
\begin{array}{ccc}
0 & 0 & \tsigma_0^{13}\\
0 & 0 & \tsigma_0^{23}\\
\tsigma_0^{13} & \tsigma_0^{23} & 0\\
\end{array}
\right) \cdot \tbfn_S - p_1 \tbfn_S - p_0 \, \tbfm_L = p_L \, \tbfm_L
$$
Since $p_0 = p_L$ on the boundary and since, in the coordinate system $(\txi_1, \txi_2, \txi_3)$, $\tbfn_S = (0,0,1)^t$, we can conclude that
\begin{equation}
\begin{array}{l}
\tsigma_0^{13}\left(\txi_1, \txi_2, 1 + \teta(\txi_1, \txi_2, \tt), \tt\right) = \tsigma_0^{23}\left(\txi_1, \txi_2, 1 + \teta(\txi_1, \txi_2, \tt), \tt\right) = 0\\
\tsigma_0^{33}\left(\txi_1, \txi_2, 1 + \teta(\txi_1, \txi_2, \tt), \tt\right) - \tp_1 = 0
\end{array}
\label{stressBC}
\end{equation}
\end{itemize}

\subsection{Component along $\boldsymbol{b_3}$}

On the component $\bfb_3$, the equations reduce to 
$$
\rho \frac{\partial u^3}{\partial t} - \left(\frac{\partial \sigma^{i3}}{\partial \xi_i} + \Gamma_{il}^i \sigma^{l3} + \Gamma_{il}^3 \sigma^{il} \right) + g^{3i} \frac{\partial p}{\partial \xi_i} = 0
$$
Then, using $g^{31}=g^{32}=0$ and $g^{33}=1$ and formulating the equations in a dimensionless form, we have
\begin{equation}
\begin{aligned}
\frac{\rho W}{T} \frac{\partial \tu^3}{\partial \tt}&
- \left(\frac{\Sigma^{13}}{R} \frac{\partial \tsigma^{13}}{\partial \txi_1} + \frac{\Sigma^{l3}}{R} \ \tCh_{1l}^1 \tsigma^{l3} + \frac{\Sigma^{1l}}{R} \ \tCh_{1l}^3 \tsigma^{1l} \right)\\
&- \left(\frac{\Sigma^{23}}{R} \frac{\partial \tsigma^{23}}{\partial \txi_2} + \frac{\Sigma^{l3}}{R} \ \tCh_{2l}^2 \tsigma^{l3} + \frac{\Sigma^{2l}}{R} \ \tCh_{2l}^3 \tsigma^{2l} \right) \\
&- \left(\frac{\Sigma^{33}}{\tau} \frac{\partial \tsigma^{33}}{\partial \txi_3} + \frac{\Sigma^{l3}}{R} \ \tCh_{3l}^3 \tsigma^{l3} + \frac{\Sigma^{3l}}{R} \  \tCh_{3l}^3 \tsigma^{3l} \right) + \frac{P}{\tau} \frac{\partial \tp}{\partial \txi_3} = 0
\end{aligned}
\end{equation}
or, once multiplied by $\frac{R^2}{\mu U}$ for getting dimensionless coefficients in front of the derivatives,
\begin{equation}
\label{b3}
\begin{aligned}
\overbrace{\frac{\rho R^2}{\mu T}}^{\footnotesize \begin{array}{c} \text{fixed to } 1\\ \text{with } T = \frac{\rho R^2}{\mu}\end{array}} \frac1{\epsilon} \frac{\partial \tu^3}{\partial \tt}&
- \left(\frac1{\epsilon} \frac{\partial \tsigma^{13}}{\partial \txi_1} 
+ \frac1{\epsilon} \ \tCh_{11}^1 \tsigma^{13} + \tCh_{11}^3 \tsigma^{11} 
+ \frac1{\epsilon} \ \tCh_{12}^1 \tsigma^{23} + \tCh_{12}^3 \tsigma^{12} 
+ \tCh_{13}^1 \tsigma^{33} + \frac1{\epsilon} \ \tCh_{13}^3 \tsigma^{13}
\right) \\
&- \left(\frac1{\epsilon} \frac{\partial \tsigma^{23}}{\partial \txi_2} 
+ \frac1{\epsilon} \ \tCh_{21}^2 \tsigma^{13} + \tCh_{21}^3 \tsigma^{21}
+ \frac1{\epsilon} \ \tCh_{22}^2 \tsigma^{23} + \tCh_{22}^3 \tsigma^{22}
+ \tCh_{23}^2 \tsigma^{33} + \frac1{\epsilon} \ \tCh_{23}^3 \tsigma^{23}
\right) \\
&- \left(\frac1{\epsilon} \frac{\partial \tsigma^{33}}{\partial \txi_3} 
+ \frac1{\epsilon} \ \tCh_{31}^3 \tsigma^{13} + \frac1{\epsilon} \  \tCh_{31}^3 \tsigma^{31}
+ \frac1{\epsilon} \ \tCh_{32}^3 \tsigma^{23} + \frac1{\epsilon} \  \tCh_{32}^3 \tsigma^{32}
+ 2 \tCh_{33}^3 \tsigma^{33}
\right)
+ \underbrace{\frac{P R}{\mu U}}_{= 1/\epsilon^2} \frac1{\epsilon} \frac{\partial \tp}{\partial \txi_3} = 0
\end{aligned}
\end{equation}
Since we have shown that all the dimensionless Christoffel symbols are at least $O(1)$ in $\epsilon$, the equations at the order $\epsilon^{-3}$ reduce to $\frac{\partial \tp_0}{\partial \txi_3} = 0$, where $\tp_0$ is the term in $\epsilon^0$ of the development in series relatively to $\epsilon$ of $\tp$.
Then, using the boundary condition on the pressure in equation (\ref{momApp}),
$$
\tp_0(\txi_1, \txi_2, \txi_3, \tt) = \tp_0(\txi_1, \txi_2, 1+\teta(\txi_1, \txi_2, \tt), \tt) = \tp_L(\xi_1, \xi_2, \tt) 
$$
At the first order in $\epsilon$, the pressure does not dependent on $\xi_3$.

At the order $\epsilon^{-2}$, the equations (\ref{b3}) reduce to $\frac{\partial \tp_1}{\partial \txi_3} = 0$, where $\tp_1$ is the term in $\epsilon^1$ of the development in series relatively to $\epsilon$ of $\tp$.
And we can conclude that $\tp_1 = 0$ since $\tp = p_L$ on the air--mucus interface.\\

\subsection{Component along $\boldsymbol{b_1}$}

As for the component $\bfb_3$, the equations on the component $\bfb_1$ reduce to
$$
\rho \frac{\partial u^1}{\partial t} - \left(\frac{\partial \sigma^{i1}}{\partial \xi_i} + \Gamma_{il}^i \sigma^{l1} + \Gamma_{il}^1 \sigma^{il} \right) + g^{1i} \frac{\partial p}{\partial \xi_i} = 0
$$
Or, in dimensionless coordinates and multiplied by $\frac{R^2}{\mu U}$,
\begin{equation}
\label{b1}
\begin{aligned}
\overbrace{\frac{\rho R^2}{\mu T}}^{=1} \frac{\partial \tu^1}{\partial \tt}
&- \left(\frac{\partial \tsigma^{11}}{\partial \txi_1} 
+ \tCh_{11}^1 \tsigma^{11} + \tCh_{11}^1 \sigma^{11}
+ \tCh_{12}^1 \tsigma^{21} + \tCh_{12}^1 \sigma^{12}
+ \frac1{\epsilon} \overbrace{\tCh_{13}^1}^{O(\epsilon)} \tsigma^{31} + \frac1{\epsilon} \overbrace{\tCh_{13}^1}^{O(\epsilon)} \sigma^{13} 
\right)&\\
&- \left(\frac{\partial \tsigma^{21}}{\partial \txi_2} 
+ \tCh_{21}^2 \tsigma^{11} + \tCh_{21}^1 \tsigma^{21}
+ \tCh_{22}^2 \tsigma^{21} + \tCh_{22}^1 \tsigma^{22}
+ \frac1{\epsilon} \overbrace{\tCh_{23}^2}^{O(\epsilon)} \tsigma^{31} + \frac1{\epsilon} \overbrace{\tCh_{23}^1}^{O(\epsilon)} \tsigma^{23} 
\right)&\\
&- \left( \frac1{\epsilon^2} \frac{\partial \tsigma^{31}}{\partial \txi_3} 
+ \overbrace{\tCh_{31}^3}^{O(\epsilon)} \tsigma^{11} + \frac1{\epsilon} \overbrace{\tCh_{31}^1}^{O(\epsilon)} \sigma^{31} 
+ \overbrace{\tCh_{32}^3}^{O(\epsilon)} \tsigma^{21} + \frac1{\epsilon} \overbrace{\tCh_{32}^1}^{O(\epsilon)} \sigma^{32} 
+ \frac1{\epsilon} \overbrace{\tCh_{33}^3}^{O(\epsilon)} \tsigma^{31} + \overbrace{\tCh_{33}^1}^{O(\epsilon)} \sigma^{33} 
\right)&\\
&+ \underbrace{\frac{P R}{\mu U}}_{= 1 / \epsilon^2} \left( \tg^{11} \frac{\partial \tp}{\partial \txi_1} 
+ \tg^{12} \frac{\partial \tp}{\partial \txi_2} \right)
= 0 &
\end{aligned}
\end{equation}
Finally, we can extract the term of the equations at the order $\epsilon^{-2}$,
$$
\frac{\partial \tsigma^{31}_0}{\partial \txi_3} = \tg^{11} \frac{\partial \tp_0}{\partial \txi_1} 
+ \tg^{12} \frac{\partial \tp_0}{\partial \txi_2}
$$
where $\tsigma^{31}_0$ is the first term  of the development in series of $\tsigma^{31}$ relatively to $\epsilon$. Since $\tp_0 = \tp_L$ does not depend on $\txi_3$, we can integrate the equation using the stress boundary conditions (equation (\ref{stressBC})) at the air--mucus interface, i.e. at $\txi_3 = 1 + \teta(\txi_1, \txi_2)$, and get
\begin{equation}
\tsigma^{31}_0(\txi_1, \txi_2, \txi_3, \tt) = - \overbrace{\left(\tg^{11} \frac{\partial \tp_L}{\partial \txi_1} + \tg^{12} \frac{\partial \tp_L}{\partial \txi_2}\right)}^{\tdpla} \left( 1 + \teta(\txi_1, \txi_2, \tt) - \txi_3 \right)
\label{sigma31}
\end{equation}
Doing a similar analysis on the component $\bfb_2$ leads to
\begin{equation}
\tsigma^{32}_0(\txi_1, \txi_2, \txi_3, \tt) = - \underbrace{\left(\tg^{21} \frac{\partial \tp_L}{\partial \txi_1} + \tg^{22} \frac{\partial \tp_L}{\partial \txi_2}\right)}_{\tdplb} \left( 1 + \teta(\txi_1, \txi_2, \tt) - \txi_3 \right)
\label{sigma31}
\end{equation}
For $i=1,2$, we define the operator $\tilde{\partial} \tilde{q}_i$ of a quantity $\tilde{q}$ as
$$
\tilde{\partial} \tilde{q}_i = \tg^{i1} \frac{\partial \tilde{q}}{\partial \txi_1} + \tg^{i2} \frac{\partial \tilde{q}}{\partial \txi_2}
$$
Then, we define the operator $\nablatx \tilde{q}$ of a quantity $\tilde{q}$ as
$$
\nablatx \tilde{q} = \tilde{\partial} \tilde{q}_1 \tilde{\bfb}_1 + \tilde{\partial} \tilde{q}_2 \tilde{\bfb}_2
$$
and $\| \nablatx \tilde{q} \| =\sqrt{\tg_{11} (\tilde{\partial} \tilde{q}_1)^2+\tg_{22}(\tilde{\partial} \tilde{q}_2)^2+2 \tg_{12} \tilde{\partial} \tilde{q}_1 \tilde{\partial} \tilde{q}_2}$.
Using these notations, we have
$$
\begin{aligned}
\tsigma^{31}_0(\txi_1, \txi_2, \txi_3, \tt) = - \tdpla \left( 1 + \teta(\txi_1, \txi_2, \tt) - \txi_3 \right)\\
\tsigma^{32}_0(\txi_1, \txi_2, \txi_3, \tt) = - \tdplb \left( 1 + \teta(\txi_1, \txi_2, \tt) - \txi_3 \right)
\end{aligned}
$$

Moreover, knowing that $\tp_1 = 0$, the terms in $\epsilon^{-1}$ in the equations (\ref{b1}) and in their equivalent on the component $\bfb_2$ lead to
$$
\begin{aligned}
\frac{\partial \tsigma^{31}_1}{\partial \txi_3} = \tg^{11} \frac{\partial \tp_1}{\partial \txi_1} + \tg^{12} \frac{\partial \tp_1}{\partial \txi_2} = 0\\
\frac{\partial \tsigma^{32}_1}{\partial \txi_3} = \tg^{21} \frac{\partial \tp_1}{\partial \txi_1} + \tg^{22} \frac{\partial \tp_1}{\partial \txi_2} = 0
\end{aligned}
$$
Consequently, both $\tsigma^{31}_1$ and $\tsigma^{32}_1$ are independent on $\txi_3$.\\

\subsection{Curvature}

The local curvature of the air--Bingham interface $\L$ parameterized as $\bfx = g(u,v,t) = f(u,v) +\tau(1+\eta(u,v,t)) \, \bfn_S(u,v)$ is
$$
\kappa(u,v) = \frac{(1+\gu^2) \gvv - 2 \gu \gv \guv + (1+\gv^2) \guu}{(1 + \gu^2 + \gv^2)^{\frac32}}
$$
Then, if we denote $\tkappa(\tu,\tv) = R \ \kappa(u,v)$ and use the previously defined dimensionless variables, then
$$
\tkappa(\tu,\tv) = \frac{(1+\tfu^2) \tfvv - 2 \tfu \tfv \tfuv + (1+\tfv^2) \tfuu}{(1 + \tfu^2 + \tfv^2)^{\frac32}} + O(\epsilon)
$$
with $f(u,v) = R \tf(\tu,\tv)$, $\tu = u / R$, $\tv = v / R$ and $\epsilon = \tau/R$.

\subsection{Model for mucus rheology}

We assume a quasi-static response of the surface tension to curvature changes and assume that the mucus behaves as a Bingham fluid as in \cite{mauroy_toward_2011, mauroy_toward_2015}.
The Bingham viscoplastic constitutive model is
\begin{equation}
\left\{
\begin{array}{ll}
\Sigma = \left( \mu + \frac{\sigma_y}{\gp} \right) \Gp & \text{for } \sigma > \sigma_y\\
\Gp = 0  & \text{for } \sigma \leq \sigma_y
\end{array}
\right.
\label{bing}
\end{equation}
with $\Gp = \left( \gp_{i,j} \right)_{i,j=1...3} = \frac12 \left( \nabla u + \left(\nabla u\right)^t \right)$.
The quantities $\sigma$ and $\gp$ are defined as $\sigma = \sqrt{\frac12 \Sigma{:}\Sigma}$ and $\gp = \sqrt{\frac12 \Gp{:}\Gp}$.

These quantities are defined in the coordinate system $(x,y,z)$.
Their expression in the coordinate system $(\xi_1, \xi_2, \xi_3)$ are obtained by using the covariant differentiation.
Thus, in the coordinates $(\xi_1, \xi_2, \xi_3)$, the derivative relatively to $\xi_j$ of the $\bfb_i$ component $u^i$ of the velocity is 
\begin{equation}
u^i_{,j} = \frac{\partial u_i}{\partial \xi_j} + \sum_{k=1}^3 \Ch_{jk}^i u^k
\label{covderiv}
\end{equation}
The dimensionless formulation of this derivative is given by the dimensionless velocities $u_i = U \tu_i$ for $i=1,2$ and $u_3 = U \epsilon \tu_3$.
Then the dimensionless formulation of the covariant derivatives are
$$
\begin{array}{ll}
u^i_{,j} = \frac{U}{R} \tu^i_{,j} & \text{for $i=1,2$ and $j=1,2$}\\
&\\
u^3_{,j} = \frac{U \epsilon}{R} \tu^3_{,j} & \text{for $j=1,2$}\\
&\\
u^i_{,3} = \frac{U}{\tau} \tu^i_{,3} & \text{for $i=1,2$}\\ 
&\\
u^3_{,3} = \frac{U \epsilon}{\tau} \tu^3_{,3} = \frac{U}{R} \tu^3_{,3} &  
\end{array}
$$
And,
\begin{equation}
\begin{array}{ll}
\tu^i_{,j} = \frac{\partial \tu^i}{\partial \txi_j} + \underbrace{\tCh_{j1}^i}_{O(1)} \tu^1 + \underbrace{\tCh_{j2}^i}_{O(1)} \tu^2 + \underbrace{\tCh_{j3}^i}_{{O(\epsilon)}} \ \epsilon \ \tu^3 = O(1) & \text{for $i=1,2$ and $j=1,2$}\\
\tu^3_{,j} = \frac{\partial \tu^3}{\partial \txi_j} + \underbrace{\tCh_{j1}^3}_{O(\epsilon)} \tu^1 + \underbrace{\tCh_{j2}^3}_{O(\epsilon)} \tu^2 + \underbrace{\tCh_{j3}^3}_{O(\epsilon)} \ \epsilon \ \tu^3 = O(1) & \text{for $j=1,2$}\\
\tu^i_{,3} = \frac1{\epsilon} \frac{\partial \tu^i}{\partial \txi_3} + \underbrace{\tCh_{31}^i}_{O(\epsilon)} \tu^1 + \underbrace{\tCh_{32}^i}_{O(\epsilon)} \tu^2 + \underbrace{\tCh_{33}^i}_{O(\epsilon)} \ \epsilon \ \tu^3 = \frac1{\epsilon} \frac{\partial u^i}{\partial \txi_3} + O(\epsilon) & \text{for $i=1,2$}\\ 
\tu^3_{,3} = \frac{\partial \tu^3}{\partial \txi_3} + \underbrace{\tCh_{31}^3}_{O(\epsilon)} \tu^1 + \underbrace{\tCh_{32}^3}_{O(\epsilon)} \tu^2 + \underbrace{\tCh_{33}^3}_{O(\epsilon)} \ \epsilon \ \tu^3 = O(1) &  
\end{array}
\label{deriv}
\end{equation}
We denote now $\gp = \frac{U}{R} \tgp$.
Then, using the dimensionless Christoffel symbols from equation (\ref{chrisnorm}) and rewriting the covariant derivatives from equation (\ref{covderiv}) in a dimensionless form, we can compute the dominant term in $\epsilon$ of the dimensionless shear rate:
$$
\tgp = \frac1{2 \epsilon} \overbrace{\tg_{33}}^{=1}\sqrt{\tg_{11}\left(\frac{\partial \tu^1}{\partial \txi_3}\right)^2 + \tg_{22}\left(\frac{\partial \tu^2}{\partial \txi_3}\right)^2 + 2 \tg_{12} \frac{\partial \tu^1}{\partial \txi_3} \frac{\partial \tu^2}{\partial \txi_3}} + O(1) = \frac1{\epsilon} E
$$
Similary, for the stress $\sigma$ with $\sigma = \frac{\mu U}{R} \tsigma$,
$$
\tsigma = \frac1{2 \epsilon} \overbrace{\tg_{33}}^{=1} \sqrt{\tg_{11} \left(\tsigma^{13}\right)^2 + \tg_{22} \left(\tsigma^{23}\right)^2 + 2 \tg_{12} \tsigma^{13} \tsigma^{23}} + O(1) = \frac1{\epsilon} T
$$
The yield condition $\sigma \geq \sigma_y$ rewrites 
$$
\tsigma \geq \frac{\sigma_y R}{\mu U} = \frac{1}{\epsilon}\frac{\sigma_y \tau}{\mu U} = \frac{1}{\epsilon} \frac{\sigma_y R^2}{\gamma \tau} = \frac{1}{\epsilon} B
$$
where $B = \frac{\sigma_y R^2}{\gamma \tau}$ is the Bingham number, that compares the yield stress to the surface tension stress. 

Under plastic conditions, i.e. $\epsilon \tsigma < B$, we have $\tgp = 0$.
Under flow conditions, i.e. when $\tsigma \geq B/\epsilon $, stress-strain relationships at the order $1/\epsilon$ are 
\begin{equation}
\left\{
\begin{aligned}
&\tsigma_0^{13} = \frac{\partial \tu_0^1}{\partial \txi_3} \left( 1 + \frac{B}{E^0} \right) = - \tdpla (1 + \teta - \txi_3)\\
&\tsigma_0^{23} = \frac{\partial \tu_0^2}{\partial \txi_3} \left( 1 + \frac{B}{E^0} \right) = - \tdplb (1 + \teta - \txi_3)\\
&E^0 = \sqrt{\tg_{11} \left(\frac{\partial \tu_0^1}{\partial \txi_3}\right)^2 + \tg_{22} \left(\frac{\partial \tu_0^2}{\partial \txi_3}\right)^2 + 2 \tg_{12} \frac{\partial \tu^1_0}{\partial \txi_3} \frac{\partial \tu^2_0}{\partial \txi_3}}
\end{aligned}
\right.
\label{firstO2}
\end{equation}
The first two equations show that $u_0^1$ and $u_0^2$ are increasing in amplitude with $\txi_3$ since their $\txi_3$ derivative is positive. 
We also have $\frac{\partial \tu_0^2}{\partial \txi_3} \ \tdpla = \frac{\partial \tu_0^1}{\partial \txi_3} \ \tdplb$ and $E^0 = |\frac{\partial \tu_0^1}{\partial \txi_3}| \ ||\nablatx \tp_L|| / |\tdpla| = |\frac{\partial \tu_0^2}{\partial \txi_3}| \ ||\nablatx \tp_L|| / |\tdplb|$, with $||\nablatx \tp_L||^2 = \tg_{11} \tdpla^2 + \tg_{22} \tdplb^2 + 2 \tg_{12} \tdpla \tdplb$.

As the stress is decreasing with $\xi_3$, if the fluid is liquid at the height $\txi_3$, then it is liquid at the height $0$.
Hence, integrating equations (\ref{firstO2}) from $0$ to $\txi_3$  and adding the boundary conditions (\ref{stressBC}) lead to
\begin{equation}
\begin{aligned}
\tu_0^1 \ = \ \frac12 \tdpla \ \txi_3 \left( \txi_3 - 2\left( 1 + \teta - \frac{B}{||\nablatx \tp_L||} \right)  \right)\\
\tu_0^2 \ = \ \frac12 \tdplb \ \txi_3 \left( \txi_3 - 2\left( 1 + \teta - \frac{B}{||\nablatx \tp_L||} \right) \right)
\end{aligned}
\label{vxvy}
\end{equation}
The stress is then
$$
\tsigma_0 = \frac1{\epsilon} \sqrt{\tg_{11} (\tsigma_0^{13})^2 + \tg_{22} (\tsigma_0^{23})^2 + 2 \tg_{12} \tsigma_0^{13} \tsigma_0^{23}} = \frac{|1+\teta-\txi_3|}{\epsilon} \times ||\nablatx \tp_L||
$$
The fluid is flowing when $\tsigma_0 \geq \frac1{\epsilon} B$ or, similarly, since $\txi_3 \leq 1+\teta$, when
$$
\txi_3 \leq \tZ(\txi_1,\txi_2) = 1 + \teta(\txi_1,\txi_2) - \frac{B}{||\nablatx \tp_L(\txi_1,\txi_2)||}
$$
$\tZ(\txi_1,\txi_2)$ is the first order yield surface.\\

Finally, the terms in $\epsilon^0$ of the velocity when $\txi_3 \leq \tZ(\txi_1,\txi_2) = 1 + \teta - \frac{B}{||\nablatx \tp_L||}$ are
\begin{equation}
\begin{aligned}
&\tu_0^1 \ = \ \frac12 \tdpla \ \txi_3 \left( \txi_3 - 2\left( 1 + \teta - \frac{B}{||\nablatx \tp_L||} \right)  \right)\\
&\tu_0^2 \ = \ \frac12 \tdplb \ \txi_3 \left( \txi_3 - 2\left( 1 + \teta - \frac{B}{||\nablatx \tp_L||} \right) \right)
\end{aligned}
\label{velo}
\end{equation}

\subsection{Dimensional fluid dynamics of the Bingham layer}
\label{dimVel}

We recall that the local coordinate system is $(\bfb_1, \bfb_2, \bfb_3)$ and that the metric tensor is $C = (\bfb_i.\bfb_j)_{ij} = (g_{ij})_{ij}$ with its inverse being $C^{-1} = (g^{ij})_{ij}$.

As for the dimensionless case, we define for $i=1,2$ the operator $\partial q_i$ of a quantity $q$ as
\begin{equation}
\partial q_i = g^{i1} \frac{\partial q}{\partial \xi_1} + g^{i2} \frac{\partial q}{\partial \xi_2}
\end{equation}
Then, we define the operator $\nablax q$ of a quantity $q$ as 
\begin{equation}
\nablax q = \partial q_1 \bfb_1 + \partial q_2 \bfb_2
\end{equation}
and 
\begin{equation}
\| \nablax q \| = \sqrt{ \nablax q \, . \,  \nablax q } = \sqrt{g_{11}(\partial q_1)^2+g_{11}(\partial q_2)^2+ 2 g_{12} \partial q_1 \partial q_2}
\end{equation}
Using these definitions, the dimensional stress is
$$
\sigma(\xi_1,\xi_2,\xi_3) = (\tau + \eta - \xi_3) \ ||\nablax p_L|| + O(\epsilon)
$$
The yield surface is located at
\begin{equation}
\label{yieldSurf}
Z(\xi_1,\xi_2) = \tau + \eta - \frac{\sigma_y}{||\nablax p_L||}
\end{equation}
In the yielded region, i.e. where $\xi_3 \leq Z(\xi_1,\xi_2)$, the velocity at height $\xi_3$ is given by
\begin{equation}
\left\{
\begin{aligned}
&u^1(\xi_1,\xi_2,\xi_3,t) = - \frac{1}{2 \mu} \dpla \ \xi_3 \left( 2 Z(\xi_1,\xi_2) - \xi_3 \right) + O\left(U \epsilon\right)\\
&u^2(\xi_1,\xi_2,\xi_3,t) = - \frac{1}{2 \mu} \dplb \ \xi_3 \left( 2 Z(\xi_1,\xi_2) - \xi_3 \right) + O\left(U \epsilon\right)\\
&u^3(\xi_1,\xi_2,\xi_3,t) = O(U \epsilon)
\end{aligned}
\right.
\end{equation}

These results indicate that the normal velocity of the Bingham fluid layer, represented by $\frac{d \bfx}{dt} = u^3(\xi_1,\xi_2,\xi_3,t) \bfb_3$ (equation (\ref{momApp})), is small relatively to the transversal velocities.

Moreover, we can exhibit a criterion on the curvature of the air--fluid interface indicating if the fluid is able to flow or not.
This condition corresponds to $Z(\xi_1,\xi_2) > 0$, or, knowing that $p_L(\xi_1,\xi_2) = -2 \gamma \kappa(\xi_1,\xi_2)$, to
\begin{equation}
\label{crit}
\|\nablax \kappa(\xi_1,\xi_2)\| < \frac{\sigma_y}{2 \gamma (\tau + \eta(\xi_1,\xi_2))}
\end{equation}

Finally, we denote $\bfu_m = (u_m^1, u_m^2, u_m^3)$ the dominant velocity averaged over the thickness of the layer written in the frame $(\xi_1, \xi_2, \xi_3)$ and:
$$
\left\{
\begin{array}{l}
u_m^1(\xi_1,\xi_2,t) = -\frac{1}{2 \mu} \dpla \hat{Z}^2(\xi_1,\xi_2)  \left( 1 - \frac{\hat{Z}(\xi_1,\xi_2) }{3 (\tau+\eta)} \right) + O(U \epsilon)\\
u_m^2(\xi_1,\xi_2,t) = -\frac{1}{2 \mu} \dplb \hat{Z}^2(\xi_1,\xi_2)  \left( 1 - \frac{\hat{Z}(\xi_1,\xi_2) }{3 (\tau+\eta)} \right) + O(U \epsilon)\\
u_m^3(\xi_1,\xi_2,t)  = O(U \epsilon)\\
\hat{Z}(\xi_1,\xi_2,t) = \max\left( 0, \tau + \eta(\xi_1,\xi_2)  - \frac{\sigma_y}{||\nablax p_L(\xi_1,\xi_2) ||} \right)
\end{array}
\right.
$$
or
\begin{equation}
\boldsymbol{u}_m(\xi_1,\xi_2,t) = -\frac{1}{2 \mu} \hat{Z}^2(\xi_1,\xi_2)  \left( 1 - \frac{\hat{Z}(\xi_1,\xi_2) }{3 (\tau+\eta)} \right) \nablax p_L + O(U\epsilon)
\end{equation}
In order to compute the integrals on the $\xi_3$ direction, we used the property that if the fluid is yielded at the height $\xi_3$, it is yielded at all the heights smaller than $\xi_3$ since the stress is decreasing with $\xi_3$. 
For $\xi_3$ larger than $Z(\xi_1,\xi_2) $, the layer is solid and its velocity is the same as the velocity at the point $(\xi_1,\xi_2,\hat{Z}(\xi_1,\xi_2))$. 

In the main text, we use the quantity $\theta(\xi_1,\xi_2) = \tau + \eta - \hat{Z}(\xi_1,\xi_2)$.

\section{Velocity of the Bingham fluid layer averaged over a bifurcation in generation $i$}
\label{meanVel}

We now assume that the thickness of the Bingham fluid layer is constant, i.e. $\eta = 0$ in the equations of the previous section. 
We consider the wall of a bifurcation $B_i$ in the generation $i$, parameterized by $\bfx = f^i(u,v)$ with $(u,v) \in \Omega_i$.
We denote $(\xi_1^i, \xi_2^i, \xi_3^i)$ as the local coordinates system in $B_i$, as defined in the Appendix \ref{coords}.
Due to the structure of our model, we know that $(\xi_1^i, \xi_2^i, \xi_3^i) = h^i \times (\xi_1^0, \xi_2^0, \xi_3^0)$.
The direction of mucocilliary clearance is represented by the unit vector $\boldsymbol{t_m}(\xi^i_1,\xi^i_2)$, as described in section \ref{mucusVel} of this document.
By definition, the component of $\boldsymbol{t_m}$ along $\xi_3$ is $0$.
We define the variation of a quantity $q$ in the direction of mucocilliary clearance with 
$$
\frac{\partial_{\xi} q}{\partial m} = \nablax q(\xi^i_1,\xi^i_2) \cdot \boldsymbol{t_m}(\xi^i_1,\xi^i_2)
$$
The Bingham layer velocity in the direction of the mucociliary clearance averaged on the whole bifurcation $B_i$ is
$$
\begin{array}{ll}
V_{m,i} & = v_{\rm{cilia}} 
-\frac1{|B_i|}\frac{1}{2 \mu} \int_{B_i\cap \{ (\xi^i_1,\xi^i_2) | Z_i(\xi^i_1,\xi^i_2) > 0 \}} \dplm(\xi^i_1,\xi^i_2) \ Z_i^2(\xi^i_1,\xi^i_2) \left( 1 - \frac{Z_i(\xi^i_1,\xi^i_2)}{3 \tau} \right) \dJ d\xi^i_1 d\xi^i_2
\end{array}
$$
The condition $Z_i(\xi^i_1,\xi^i_2) > 0$ 
can be reformulated in $(\xi^0_1, \xi^0_2, \xi^0_3)$ using ${Z_0}(\xi^0_1,\xi^0_2)$ and $i$.
It becomes ${Z_0}(\xi^0_1,\xi^0_2) > \tau \left( 1 - \frac1{h^{2i} }\right)$. 
Moreover $Z_i(\xi^i_1,\xi^i_2) = \tau - h^{2i} (\tau - Z_0(\xi_1^0,\xi_2^0))$.

Now, we recall that $\nablax p_{L,i}(\xi^i_1,\xi^i_2) = - 2 \gamma \nablax \kappa_i(\xi^i_1,\xi^i_2) = - \frac{2 \gamma}{h^{2i}} \nablax \kappa_0 (\xi^0_1,\xi^0_2)$ and we denote
$$
D_i = B_0 \cap \left\{ (\xi^0_1,\xi^0_2) \ | \ ||\nablax \kappa_0(\xi^0_1,\xi^0_2) || > \frac{\sigma_y}{2 \tau \gamma} h^{2i} \right\}
$$
Then, we can write the previous equation on $V_{m,i}$ using powers of $\left(\frac{\sigma_y h^{2i}}{2 \tau \gamma} \frac1{|| \nablax \kappa_{0}(\xi^0_1,\xi^0_2) ||}\right)$,
\begin{equation}
\begin{array}{ll}
V_{m,i} = & v_{\rm{cilia}}\\
& + \frac1{h^{2i}}\frac{2 \gamma \tau^2}{3 \mu |B_0|} \int_{D_i} \ \dkm(\xi^0_1,\xi^0_2) \left(\frac{\sigma_y h^{2i}}{2 \tau \gamma} \frac1{|| \nablax \kappa_{0}(\xi^0_1,\xi^0_2) ||}\right)^0 \ \dJz d\xi^0_1 d\xi^0_2\\
\\
& - \frac1{h^{2i}} \frac{\gamma \tau^2}{\mu |B_0|} \int_{D_i} \ \dkm(\xi^0_1,\xi^0_2) \left(\frac{\sigma_y h^{2i}}{2 \tau \gamma} \frac1{|| \nablax \kappa_{0}(\xi^0_1,\xi^0_2) ||}\right)^1 \ \dJz d\xi^0_1 d\xi^0_2\\
\\
& + \frac1{h^{2i}} \frac{\gamma \tau^2}{3 \mu |B_0|} \int_{D_i} \ \dkm(\xi^0_1,\xi^0_2) \left(\frac{\sigma_y h^{2i}}{2 \tau \gamma} \frac1{|| \nablax \kappa_{0}(\xi^0_1,\xi^0_2) ||}\right)^3 \ \dJz d\xi^0_1 d\xi^0_2\\
&\\
V_{m,i} = & v_{\rm{cilia}} + \frac1{h^{2i}}\frac{2 \gamma \tau^2}{3 \mu |B_0|} \int_{B_0 \cap \{ (\xi^0_1,\xi^0_2) \ | \ X_i(\xi^0_1,\xi^0_2) < 1 \}} \ \dkm(\xi^0_1,\xi^0_2) \times\\
& \hspace{7cm}
\left(
1 - \frac32 X_i(\xi^0_1,\xi^0_2)
+ \frac12 X_i(\xi^0_1,\xi^0_2)^3 
\right) \ \dJz d\xi^0_i d\xi^0_2
\end{array}
\label{totalVel}
\end{equation}
with $X_i(\xi^0_1,\xi^0_2) = \frac{\sigma_y}{2 \tau \gamma} \frac{h^{2i}}{||\nablax \kappa_0(\xi^0_1,\xi^0_2) ||}$. 
If $X_i$ is close to $1$, then $1 - \frac32 X_i(\xi^0_1,\xi^0_2) + \frac12 X_i(\xi^0_1,\xi^0_2)^3$ is small and the relative contribution to the whole integral is small. 
On the contrary, if $X_i$ is small, then the relative contribution to the integral is maximal. 
Consequently, the integral is dominated by the contribution of the regions where $X_i$ is small, i.e. where the curvature gradient is large.

%%%%%%%%%%%%%%%%%%%%%%%%%%%%%%%%%%%%%%%%%%%%%%%%%%%%%%%%%%%%%%%%%%%%%%

\section{Numerical simulations of the motion of a layer of a Bingham fluid on the wall of an airway tree}
\label{sim}

\subsection{Geometry}
\label{curvSmooth}
The geometry of the three-generation airway tree is based on typical size ratios measured in the lung \cite{tawhai_ct-based_2004}. 
The root branch diameter is $1 \ mm$, and the branch size decreases at each bifurcation with the ratio $\left(\frac12\right)^{\frac13}$. 
The ratio of length to diameter is $3$. 
Two successive branching planes form an angle of $90$ degrees with each other. 
The CAD geometry for GMSH is automatically built using Octave, and STL surface meshes are generated using GMSH \cite{geuzaine_gmsh:_2009}.
Visual details are provided in Fig. \ref{mesh}.

\begin{figure}[h!]
\centering 
\includegraphics[height=4cm]{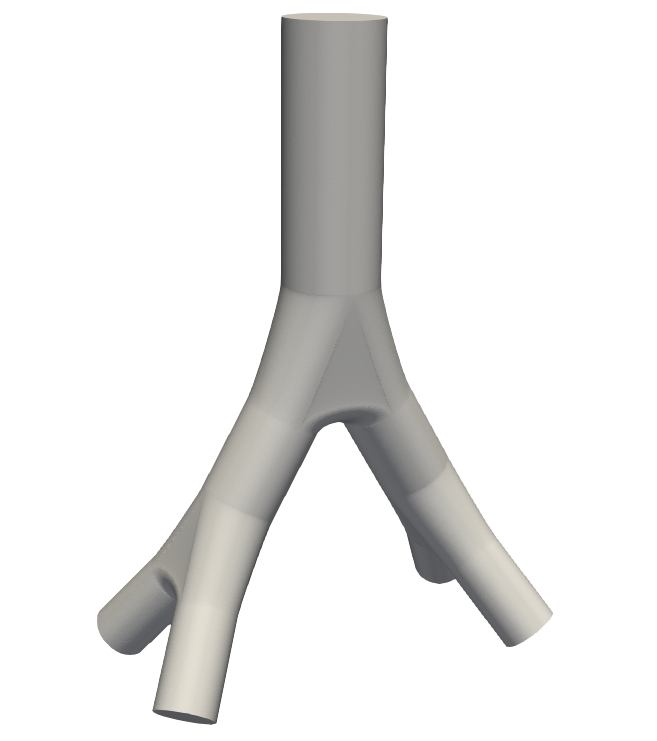}
\includegraphics[height=4cm]{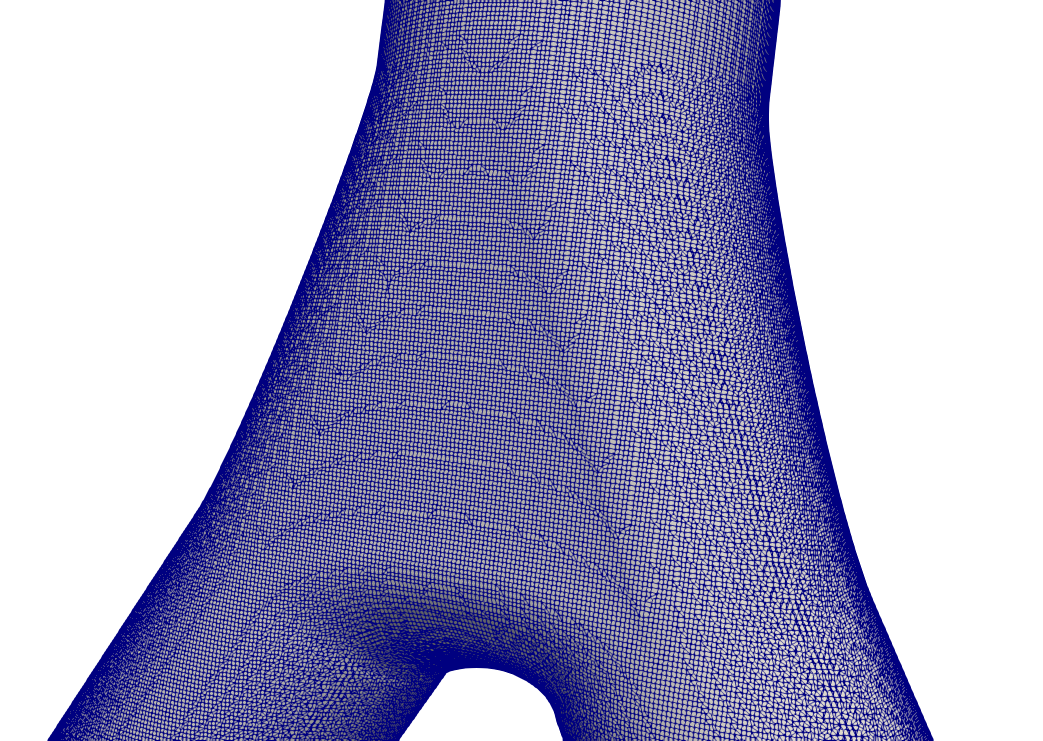}
\caption{The geometry and its surface mesh used in the simulations. 
}
\label{mesh}
\end{figure}

The curvature is computed from the surface divergence $k_0 = \Div_S (\bfn)$ of the inwards normals using boundary finite elements. 

A (reasonably) crude mesh provides a good characterization of the main features of a bifurcation but affects the quality of the variables computed using the finite elements method.
On the contrary, an extremely fine mesh offers high-quality estimations with the finite elements method but introduces noise in the curvature, which is not meaningful to our approach. 
Thus, the curvature is smoothed to allow the use of a mesh fine enough for the finite elements method while capturing the main geometrical features of the bifurcation without noise.
The curvature is smoothed using a technique from image analysis based on the heat equation~\cite{guichard_review_2002}.
The method involves "applying" the heat equation to the divergence of the normals to the bifurcations $k_0 = \Div_S (\bfn)$.
More precisely, $k_0$ corresponds to the initial state in the following partial differential equation:
$$
\left\{
\begin{aligned}
&\frac{\partial k}{\partial e}(\xi,e) - D \bigtriangleup_S k (\xi,e) = 0 \text{ for } (\xi,e) \in \Omega \times ]0,1]\\
&k(\xi,0) = k_0(\xi)
\end{aligned}
\right.
$$
Then, the curvature used in our work is the field $k$ taken at the time $e=1$, i.e. $\kappa = k(.,1)$.
The resulting smoothing of the curvature corresponds to a convolution of the initial curvature field with a bi-dimensional Gaussian kernel 
$$
K(\xi) = \frac{1}{2\pi \std^2 } e^{-\frac{|\xi|^2}{2 \std^2}}
$$ 
with a standard deviation $\std = \sqrt{2 D}$.

\begin{figure}[t!h]
\centering
\includegraphics[height=6cm]{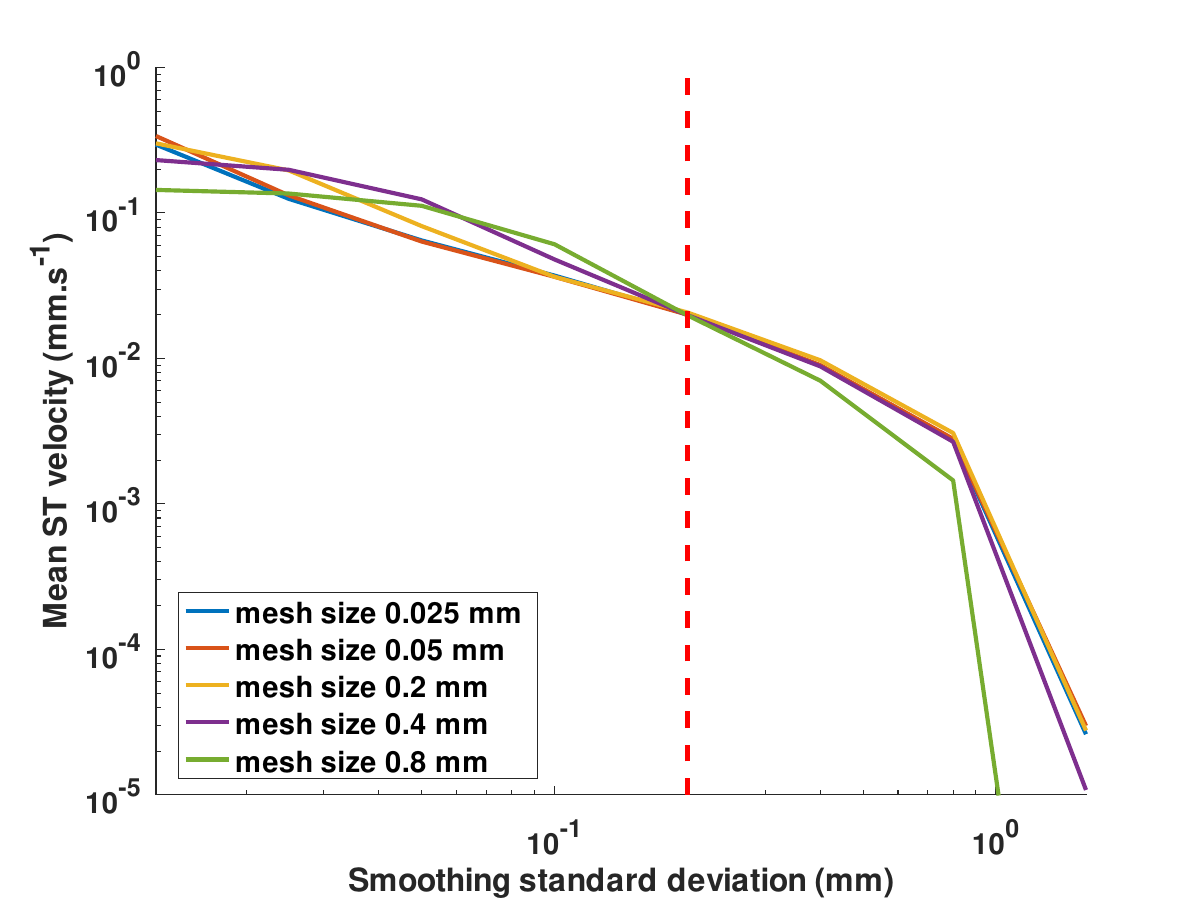}
\caption{Sensitivity of the mean Bingham fluid velocity (log-log) in the bifurcation relative to curvature smoothing (x-axis) and mesh refinement (colored curves).
The finer the mesh, the more precise the quality of the finite elements method, but also the more sensitive the curvature is to the mesh.
If the standard deviation of smoothing is too small, the fluid velocity is influenced by mesh specificities; conversely, if it is too large, the features of the bifurcation are lost.
The chosen degree of smoothing is $\std = 0.2$ mm.
A smaller value introduces artefacts due to the discretization into triangles of the bifurcation surface, while a larger value leads to over-smoothing, hiding the main geometrical features of the bifurcation.
The chosen mesh size is $0.05$ mm, corresponding to $186\, 594$ triangles for meshing the bifurcation surface.
}
\label{smoothMesh}
\end{figure}

We tested how the smoothing affects the mean Bingham fluid velocity in the bifurcation as a function of the mesh refinement, as shown in Fig. \ref{smoothMesh}.
The degree of smoothing was then determined by the value for which the velocity was the closest for all the meshes tested, indicating that the result does not depend on the mesh size.
The mesh size was fixed at $0.05$ mm, and the standard deviation of the smoothing was set to $\std = 0.2$ mm, corresponding to a diffusive coefficient $D = 2 \ 10^{-8}$ m$^2$.s$^{-1}$. 
The resulting smoothed curvature field is then used as the input for the model computations.

\subsection{Numerical simulations}

To study the properties of a thin layer of Bingham fluid in a 3D geometry, we used boundary finite elements within Comsol Multiphysics 3.5a.

The healthy layer thickness chosen in our work corresponds to the most frequent reported mean value in the literature: $\tau = 10 \ \mu$m \cite{karamaoun_new_2018}.
Several other thickness values have been simulated to mimic pathological mucus layer, up to $\tau = 150 \ \mu$m.

We estimated the characteristic size of the domain $R$ using the airway radii.
The thickness can be considered small relative to the curvature radius in most generations of the tree.
We indicated in the results when this hypothesis breaks.
We use the results from the lubrication theory of a Bingham fluid to estimate the main component of the thin Bingham fluid layer velocity.

The embedded capability of Comsol Multiphysics 3.5a was used to compute the tangential and normal vectors of a surface.
Moreover, these vectors define an orthonormal local basis, and the metric tensors $g^{(i,j)}$ and $g_{(i,j)}$ are equal to the identity matrix. 
As a consequence, the dominant velocities of the Bingham layer averaged over its thickness are expressed as
$$
\left\{
\begin{array}{l}
u_m^1(\xi_1,\xi_2) = -\frac{1}{2 \mu} \frac{\partial p_L}{\partial \xi_1} \hat{Z}^2(\xi_1,\xi_2)  \left( 1 - \frac{\hat{Z}(\xi_1,\xi_2) }{3 (\tau+\eta)} \right) + O(U \epsilon)\\
u_m^2(\xi_1,\xi_2) = -\frac{1}{2 \mu} \frac{\partial p_L}{\partial \xi_2} \hat{Z}^2(\xi_1,\xi_2)  \left( 1 - \frac{\hat{Z}(\xi_1,\xi_2) }{3 (\tau+\eta)} \right) + O(U \epsilon)\\
u_m^3(\xi_1,\xi_2)  = O(U \epsilon)\\
\hat{Z}(x,y) = \max\left( 0, \tau + \eta(\xi_1,\xi_2)  - \frac{\sigma_y}{||\nablax p_L(\xi_1,\xi_2) ||} \right)
\end{array}
\right.
$$

We use the embedded surface derivatives in Comsol Multiphysics to compute the surface divergence of the normal $\boldsymbol{n}$ to the surface.
The mean curvature of the airway wall surface is then calculated as $\kappa = \frac12 \Div_{\xi}(\boldsymbol{n})$.
To avoid a noisy curvature resulting from the meshing of the surface, the computed curvature is locally smoothed using a kernel, the width of which is determined in section \ref{curvSmooth} of this Supplemental Materials.  

We assume that the air pressure in the airways is $0$.
Then, the Laplace pressure is computed as $p_L = - 2 \gamma \kappa$, where $\gamma$ is the surface tension.

Finally, we again use the embedded surface derivatives in Comsol Multiphysics to compute the derivatives of $p_L$, $\frac{\partial p_L}{\partial \xi_1}$, and $\frac{\partial p_L}{\partial \xi_1}$ along the tangential directions to the airway walls. 

\section{Estimating the orientation of cilia velocity}
\label{mucusVel}

The mucociliary motion of mucus was simulated using a velocity at the airways wall with an amplitude $v_{\rm{cilia}} = 50 \ \mu m.s^{-1}$. 
The directions of the mucocilliary motion were determined by the directions of the gradient of a Laplacian field $L$, with the following boundary conditions: $L=1$ at the opening of the largest airway of the bifurcation, and $L=0$ at the opening of the smallest airways of the bifurcation. 
No $L$ flow was permitted through the wall of the tree. 
The wall gradient of such a field is smooth, tangent to the wall, and parallel to the centerlines of the tree.
We assumed that the velocity induced on mucus by mucocilliary transport is 
$$
\boldsymbol{v}_{\rm{cilia}} = v_{\rm{cilia}} \times \frac{\nabla L}{||\nabla L||}
$$
\begin{figure}[h!]
\centering 
\includegraphics[height=6cm]{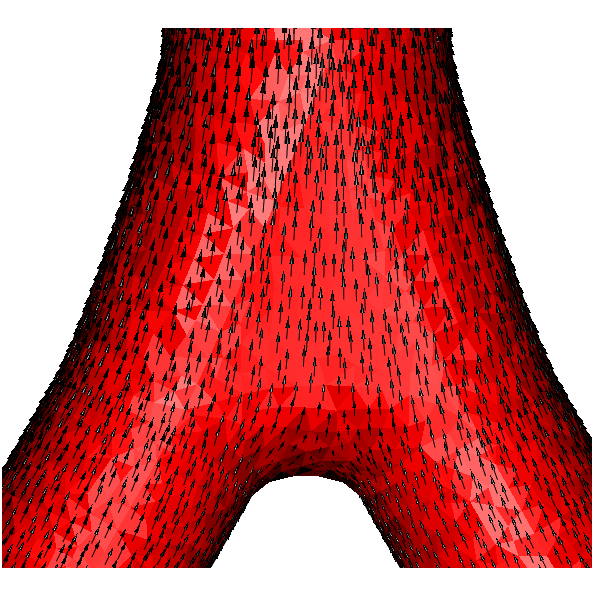}
\caption{Details of a bifurcation to show the direction of the motion of mucus due to cilia, predicted by our model based on the gradient of a Laplacian field.}
\label{mesh}
\end{figure}
Another mathod for estimating the velocity field induced by mucocilliary clearance is proposed in \cite{manolidis_macroscopic_2016}.
The properties of the field obtained by our method and in \cite{manolidis_macroscopic_2016} are very close. 
The method proposed here allows for computing a velocity wall field without explicitly computing the centerlines of the tree, which can be useful for complex geometries.

As discussed in \cite{karamaoun_new_2018}, assuming a generation-independent velocity amplitude for the mucus layer is not compatible with a constant mucus layer thickness throughout the tree. 
Indeed, considering a branch in generation $i$ with radius $r_i$ that bifurcates into two branches in generation $i+1$ with radii $r_{i+1} = h r_i$, we can relate the mucus layer thicknesses $\tau_i$ and $\tau_{i+1}$ between the two generations:
$$
\underbrace{2 \pi r_i \ \tau_i \ v_{\rm{cilia}}}_{\text{outflow of branch $i$}} = \underbrace{2 \times 2 \pi r_{i+1} \ \tau_{i+1} \ v_{\rm{cilia}}}_{\text{outflow of branches $i+1$}} \ \longrightarrow \ \tau_i = 2 h \ \tau_{i+1} \simeq 1.59 \ \tau_{i+1}
$$
Thus, the small differences in term of mucus layer thicknesses between the bronchial generations likely result from regulation by other mechanisms, which are not well described as of today \cite{karamaoun_new_2018}. 
The way mucociliary clearance is simulated in this study does not account for such potential other regulatory mechanisms.

\bibliographystyle{abbrv}      % mathematics and physical sciences
%\bibliographystyle{unsrtnat}
%\bibliography{bibli_2023}  